\documentclass[prd,twocolumn,superscriptaddress,amsmath,amssymb]{revtex4-1} 

\usepackage{bm}
\usepackage{dcolumn}

\begin{document}

\title{Residual Symmetries of the Gravitational Field}
\thanks{We dedicate this modest contribution to Albert Einstein in the
centennial of General Relativity, who opened the door to an unprecedent
understanding of the gravitational field.}

\author{Eloy Ay\'on-Beato}
\email{ayon-beato-at-fis.cinvestav.mx}
\affiliation{Departamento~de~F\'{\i}sica,~CINVESTAV--IPN,%
~Apdo.~Postal~14--740,~07000,~M\'exico~D.F.,~M\'exico}
\affiliation{Instituto~de~Ciencias~F\'isicas~y~Matem\'aticas,%
~Universidad~Austral~de~Chile,~Valdivia,~Chile}

\author{Gerardo Vel\'{a}zquez-Rodr\'{\i}guez}
\email{gvelazquez-at-fis.cinvestav.mx}
\affiliation{Departamento~de~F\'{\i}sica,~CINVESTAV--IPN,%
~Apdo.~Postal~14--740,~07000,~M\'exico~D.F.,~M\'exico}
\affiliation{Instituto~de~Ciencias~F\'isicas~y~Matem\'aticas,%
~Universidad~Austral~de~Chile,~Valdivia,~Chile}

\begin{abstract}
We develop a geometric criterion that unambiguously characterizes the
residual symmetries of a gravitational \emph{Ansatz}. It also provides a
systematic and effective computational procedure for finding all the
residual symmetries of any gravitational \emph{Ansatz}. We apply the criterion to
several examples starting with the Collinson \emph{Ansatz} for circular stationary
axisymmetric spacetimes. We reproduce the residual symmetries already known
for this \emph{Ansatz} including their conformal symmetry, for which we identify
the corresponding infinite generators spanning the two related copies of
the Witt algebra. We also consider the noncircular generalization of this
\emph{Ansatz} and show how the noncircular contributions on the one hand break the
conformal invariance and on the other hand enhance the standard
translations symmetries of the circular Killing vectors to
supertranslations depending on the direction along which the circularity is
lost. As another application of the method, the well-known relation
defining conjugate gravitational potentials introduced by Chandrasekhar,
that makes possible the derivation of the Kerr black hole from a trivial
solution of the Ernst equations, is deduced as a special point of the
general residual symmetry of the Papapetrou \emph{Ansatz}. In this derivation we
emphasize how the election of Weyl coordinates, which determines the
Papapetrou \emph{Ansatz}, breaks also the conformal freedom of the stationary
axisymmetric spacetimes. Additionally, we study AdS waves for any dimension
generalizing the residual symmetries already known for lower dimensions and
exhibiting a very complex infinite-dimensional Lie algebra containing three
families: two of them span the semidirect sum of the Witt algebra and
scalar supertranslations and the third generates vector supertranslations.
Independently of this complexity we manage to comprehend the true meaning
of the infinite connected group as the precise diffeomorphisms subgroup
allowing to locally deform the AdS background into AdS waves.
\end{abstract}

\pacs{}

\maketitle


\section{\label{sec:Int}Introduction}

A frequent scenario in the now centennial theory of general relativity is
that solving Einstein equations can lead to integration constants and even
integration functions that turn out to be redundant, i.e.\ they can be
eliminated by appropriate coordinate transformations and are not related to
the truly dynamical content of the theory. This is due to the fact that after
realizing the original symmetries of the physical configuration under study
through a chosen \emph{Ansatz}, a sort of gauge freedom can remains, defining a
residual symmetry on the defining variables of the \emph{Ansatz}. Consequently,
these eliminations can be expected previous to the integration process by
virtue of its kinematical character. In order to illustrate the point, let us
consider as an example the Collinson \emph{Ansatz} for a circular stationary
axisymmetric spacetime \cite{Collinson:1976,Garcia:2002gj,AyonBeato:2005yx}
\begin{equation}\label{eq:coll}
\bm{\mathrm{d}s^2}=e^{-2Q}\left(
-\frac{(\bm{\mathrm{d}\tau}+a\bm{\mathrm{d}\sigma})
(\bm{\mathrm{d}\tau}-b\bm{\mathrm{d}\sigma})}{a+b}
+e^{-2P}\bm{\mathrm{d}z\mathrm{d}\bar{z}}\right),
\end{equation}
where the structural functions $a$, $b$, $P$ and $Q$ only depend on the
complex coordinate $z$ and its complex conjugate $\bar{z}$.

If one performs
the following diffeomorphism
\begin{equation}\label{eq:diff}
(\tau,\sigma, z)\,\mapsto\,
(\tilde{\tau}=\tau+\varepsilon\sigma, \tilde{\sigma}=\sigma, \tilde{z}=z),
\end{equation}
it can be compensated for the next redefinition of the structural functions
\begin{equation}\label{eq:comp}
(a,b,P,Q)\,\mapsto\,
(\tilde{a}=a-\varepsilon, \tilde{b}=b+\varepsilon, \tilde{P}=P, \tilde{Q}=Q),
\end{equation}
which means that in terms of the new variables the metric is form invariant
\begin{equation}\label{eq:collb}
\bm{\mathrm{d}s^2}=e^{-2\tilde{Q}}\left(
-\frac{(\bm{\mathrm{d}\tilde{\tau}}+\tilde{a}\bm{\mathrm{d}\tilde{\sigma}})
(\bm{\mathrm{d}\tilde{\tau}}-\tilde{b}\bm{\mathrm{d}\tilde{\sigma}})}
{\tilde{a}+\tilde{b}}+e^{-2\tilde{P}}
\bm{\mathrm{d}\tilde{z}\mathrm{d}\bar{\tilde{z}}}\right).
\end{equation}
This fact implies that if a solution for the functions $a$ and $b$ is found
such that both contain homogeneous contributions differing only in sign, as
the ones exhibited in the redefinitions \eqref{eq:comp}, these contributions
can be eliminated with the previous diffeomorphism \eqref{eq:diff}, which is
just a simple rotation in the $(\tau,\sigma)$-plane.

The above is a concrete example of a residual symmetry of the Collinson
\emph{Ansatz}, and it is natural to wonder whether there are more such symmetries
and obviously how to find them. Unfortunately, these kind of symmetries are
usually discovered by the large use of experience and having an eye for that
is crucial. Hence, it is fundamental to understand if there exists a general
way to characterize residual symmetries that additionally provide a
systematic way to find all of them. The main objective of this work is
precisely to propose a criterion which allows to accomplish these goals for
any metric \emph{Ansatz}.

More concretely, suppose you have a metric whose components $g_{\alpha\beta}$
explicitly depend on some of the coordinates $x^\mu$ and its remaining
dependence implicitly occurs through some \emph{a priori} unknown functions
$u^J=u^J(x^{\mu})$, giving rise to a concrete metric \emph{Ansatz}
\begin{equation}\label{eq:Ansatz}
\bm{\mathrm{d}s^2}=g_{\alpha\beta}(x^{\mu},u^J)
\bm{\mathrm{d}x}^{\alpha}\bm{\mathrm{d}x}^{\beta}.
\end{equation}
We are interested in the following question: what are the continuous
transformations that preserve the form of this \emph{Ansatz}? Namely, coordinate
transformations
\begin{subequations}\label{eq:Lie_pt}
\begin{equation}\label{eq:Lie_ptc}
x^\alpha\mapsto\tilde{x}^\alpha=\tilde{x}^\alpha(x^{\mu},u^J;\varepsilon),
\end{equation}
and redefinitions of the \emph{Ansatz} functions
\begin{equation}\label{eq:Lie_ptf}
u^{I}\mapsto\tilde{u}^{I}=\tilde{u}^{I}(x^{\mu},u^J;\varepsilon),
\end{equation}
\end{subequations}
depending on at least one parameter $\varepsilon$, such that they satisfy
\begin{align}\label{eq:cond1}
\bm{\mathrm{d}s^2}&=g_{\alpha\beta}(x^{\mu},u^J)
\bm{\mathrm{d}x}^{\alpha}\bm{\mathrm{d}x}^{\beta}\nonumber\\
&=g_{\alpha\beta}(\tilde{x}^{\mu},\tilde{u}^J)
\bm{\mathrm{d}\tilde{x}}^{\alpha}\bm{\mathrm{d}\tilde{x}}^{\beta}
=\bm{\mathrm{d}\tilde{s}^2},
\end{align}
where the latter expression means that the dependence of the gravitational
potentials $g_{\alpha\beta}$ in terms of the new quantities must be exactly
the same than in the original variables. We emphasize that, in order to cover
the more general situation, it is convenient to consider mixed
transformations where the coordinates $x^\mu$ and the unknown functions $u^J$
are handled as independent variables, i.e.\ the transformations
(\ref{eq:Lie_pt}) are just point to point maps in a space built of spacetime
coordinates together with the \emph{Ansatz} functions. The advantage of dealing with
a parametric family lies in the fact that we can assume that
(\ref{eq:Lie_pt}) form a one-parameter group of point transformations or,
equivalently, a Lie point transformation and apply the machinery of
continuous symmetry groups originally introduced by Sophus Lie for the
systematic study of differential equations \cite{StH,OlP,SoL,Blu}.

The Lie point symmetries of vacuum Einstein equations are well understood
\cite{Ibragimov,Marchildon:1997xc}, as well as their local Lie-B\"{a}cklund
generalizations \cite{Torre:1993jm,Capovilla:1993cs,Anderson:1994eg}. Even it
is known how to implement non-local Lie-B\"{a}cklund generalizations in
presence of Killing vectors, see \cite{Stephani2003} and references therein.
However, for the case of configurations enjoying a given symmetry as the ones
in which we are interested, also called group invariant solutions, no
systematic study is known about their residual symmetries. An exception is
Ref.~\cite{Anderson:1999cn} where the Lie symmetry reduction method is
reexamined in the context of General Relativity in comparison with the
standard derivations of exact solution in this field. Residual symmetries
plays an important role in this construction, they were termed the
\emph{automorphism group} in this reference, where the expression
\emph{residual group} was reserved for the symmetries of the reduced field
equations. Another exception is Ref.~\cite{Papadopoulos:2011rk} which focuses
the study of residual symmetries to the case of homogeneous spaces.

The work is organized as follows. In Sec.~\ref{sec:Crit} we establish the
theoretical framework used in order to derive the criterion to study the
residual symmetries of a generic metric \emph{Ansatz}. With the criterion at hand,
in Sec.~\ref{sec:Coll} we return to the subject of the residual symmetries of
the Collinson \emph{Ansatz} (\ref{eq:coll}). After solving the restrictions imposed
by the criterion we consistently found the whole class already exhibited by
Collinson himself \cite{Collinson:1976}. In Sec.~\ref{sec:Spher} we treat the
well-known case of spherically symmetric spacetimes in its static version as
well as the time-dependent one and we emphasize its similarities and
differences. We investigate the so-called anti–de Sitter (AdS) waves for any dimension in
Sec.~\ref{sec:AdS}, we generalize the residual symmetries of the
lower-dimensional cases presented in
Refs.~\cite{Siklos:1985,AyonBeato:2005qq}. The renowned Papapetrou \emph{Ansatz}
\cite{Papapetrou:1953zz} is studied in Sec.~\ref{sec:Papa}. The knowledge of
its more general residual symmetry transformation allows us to find as a
particular limiting case the conjugate transformation between the Ernst
potentials introduced by Chandrasekhar \cite{HeM}. We remark this conjugate
transformation is the one allowing to relate a trivial but unphysical
solution of the Ernst equations \cite{Ernst:1967wx} to the celebrated Kerr
black hole \cite{Kerr:1963ud}. Finally, we explore the noncircular
generalization of the Collinson \emph{Ansatz} \cite{AyonBeato:2005yx} in
Sec.~\ref{sec:noncircColl}. In this case, the noncircular contributions
change dramatically the nature of the residual symmetries of the \emph{Ansatz} if
they are compared to those of the original Collinson metric. The detailed
calculations showing how to solve the proposed criterion for each case are
included in separate appendixes in order to emphasize the simplicity of the
method.

\section{\label{sec:Crit}The criterion}

In this section we will provide a geometrical characterization of residual
symmetries. The starting point is to think of coordinates and structural
functions $\grave{a}$ \emph{la} Lie, i.e.\ all handled as independent variables in
an abstract space (we return to this point later). After that, we consider
the infinitesimal version of the one-parameter Lie point transformations
defining the residual symmetries (\ref{eq:Lie_pt})
\begin{subequations}\label{eq:trans1}
\begin{align}
\tilde{x}^{\alpha}&=x^\alpha+\varepsilon\xi^\alpha(x^{\mu},u^{J})+\cdots,\\
\tilde{u}^{I}&=u^{I}+\varepsilon\eta^{I}(x^{\mu},u^{J})+\cdots,
\end{align}
\end{subequations}
which give rise to the generator
\begin{equation}\label{eq:gengen}
\bm{X}=\xi^{\alpha}(x^{\mu},u^{J})\bm{\partial}_{\alpha}
+\eta^{I}(x^{\mu},u^{J})\bm{\partial}_{I},
\end{equation}
with components defined as usual
\begin{subequations}\label{eq:compts}
\begin{align}
\xi^{\alpha}(x^{\mu},u^{I})&\equiv{
\frac{\partial\tilde{x}^\alpha}{\partial\varepsilon}
\hspace{0.05in}\vline}_{\hspace{0.05in}\varepsilon=0},
\label{eq:spacecompts}\\
\eta^{I}(x^{\mu},u^{J})&\equiv{
\frac{\partial\tilde{u}^{I}}{\partial\varepsilon}
\hspace{0.05in}\vline}_{\hspace{0.05in}\varepsilon=0}.
\label{eq:structcompts}
\end{align}
\end{subequations}
For example, it is easy to see that the residual symmetry of the Collinson
\emph{Ansatz} defined by the transformations \eqref{eq:diff} and \eqref{eq:comp}
coincides with their infinitesimal version, giving as generator
\begin{equation}\label{eq:diffcomp_gen}
\bm{X}=\sigma\bm{\partial_\tau}-\bm{\partial_a}+\bm{\partial_b}.
\end{equation}

We are now ready to explore the infinitesimal consequences of the
form-invariant condition (\ref{eq:cond1})
\begin{align}
\bm{\mathrm{d}\tilde{s}^2}={}&g_{\alpha\beta}(x^\mu+\varepsilon\xi^\mu+\cdots,
u^J+\varepsilon\eta^J+\cdots)\frac{}{}\nonumber\\
&\times\bm{\mathrm{d}}(\bm{x}^\alpha+\varepsilon\bm{\xi}^\alpha+\cdots)
\bm{\mathrm{d}}(\bm{x}^\beta+\varepsilon\bm{\xi}^\beta+\cdots)\frac{}{}
\nonumber\\
={}&\bm{\mathrm{d}s^2}
+\varepsilon\bigl[\left(\xi^{\mu}\partial_{\mu}g_{\alpha\beta}
+2g_{\mu\alpha}\partial_{\beta}\xi^{\mu}
+\eta^{I}\partial_{I}g_{\alpha\beta}\right)
\!\bm{\mathrm{d}x}^{\alpha}\bm{\mathrm{d}x}^{\beta}
\frac{}{}\nonumber\\
&\qquad\qquad
+2g_{\alpha\beta}\partial_{I}\xi^{\beta}
\bm{\mathrm{d}x}^{\alpha}\bm{\mathrm{d}u}^{I}
\bigr]+\cdots,\frac{}{}
\label{eq:invcond}
\end{align}
i.e.\ keeping the form of the metric requires necessarily the following
conditions on the infinitesimal generators
\begin{subequations}\label{eq:critinf}
\begin{align}
\xi^{\mu}\partial_{\mu}g_{\alpha\beta}
+2g_{\mu(\alpha}\partial_{\beta)}\xi^{\mu}
&=-\eta^{I}\partial_{I}g_{\alpha\beta},
\label{eq:crit2}\\
\partial_{I}\xi^{\beta}&=0.\label{eq:crit1}
\end{align}
\end{subequations}
These conditions are interpreted as follows. The condition \eqref{eq:crit1}
establishes that the components of the generator along spacetime depend only
on the spacetime coordinates, i.e.\ these components represent genuine
one-parameter diffeomorphisms of spacetime generated by the vector field
$\bm{\xi}=\xi^{\mu}(x^\alpha)\bm{\partial}_{\mu}$. At the same time, a
one-parameter diffeomorphism change the metric by the corresponding Lie
derivative, $\bm{\pounds_{\xi}g}_{\mu\nu}$, which is just what appears to the
left hand side of condition \eqref{eq:crit2}. Hence, the fact that the right
hand side of condition \eqref{eq:crit2} is minus the directional derivative
of the metric along the components of the generator associated with the
structural functions,
$\bm{\eta}(g_{\mu\nu})=\eta^I\bm{\partial}_Ig_{\mu\nu}$, only remarks that
the above spacetime diffeomorphisms are compensated by appropriate
redefinitions of the structural functions:
$\bm{\pounds_{\xi}g}_{\mu\nu}=-\bm{\eta}(g_{\mu\nu})$. These conditions are
entirely compatible with the given definition of residual symmetry and are
obtained by an elementary infinitesimal reasoning.

For example, it is easy to check that the generator \eqref{eq:diffcomp_gen}
indeed satisfy the conditions \eqref{eq:critinf} once they are evaluated in
the Collinson \emph{Ansatz}. Regarding the question posed at the introduction on
whether there are more residual symmetries, we emphasize that for any given
fixed gravitational background the conditions \eqref{eq:critinf} are an
over-determined system of linear partial differential equations for the
components of the generator as functions of the coordinates and the
structural functions; the general solution for such systems can always be
found \cite{OlP}.

Before embarking on the expedition to explore the consequence of the obtained
conditions on concrete gravitational backgrounds, we would like to highlight
that the Lie-point transformation \eqref{eq:Lie_pt} describing residual
symmetries, its infinitesimal version \eqref{eq:trans1} with generator
\eqref{eq:gengen}, and the resulting conditions \eqref{eq:critinf} have all a
more geometrical and intuitive interpretation if we consider an abstract
space with coordinates
\begin{equation}\label{eq:z^A}
z^{A}=\left(x^{\mu},u^{I}\right).
\end{equation}
In applications to differential equations is common the use of a prolonged
version of this space which includes also the derivatives of structural
functions as coordinates,
$z^{A}=\left(x^{\mu},u^{I},u^{I}_{\mu_{1}},u^{I}_{\mu_{1}\mu_{2}},
\ldots,u^{I}_{\mu_{1}\ldots\mu_{n}}\right)$, and it is formally known in
mathematics as the $n$-th \emph{jet space} \cite{OlP}, where
$u^{I}_{\mu_{1}\ldots\mu_{n}}$ denotes the $n$-th derivative of the
structural function $u^{I}$ with respect to spacetime coordinates $x^{\mu}$.
In our case, the metric does not depend on derivatives, therefore, it is
sufficient to consider the $0$th jet space or simply jet
space.\footnote{There are some special examples of metric \emph{Ansatz} which also
depend on the derivatives of the structural functions, they deserve a
separate attention \cite{Ayon-Beato:2015awd}.} The residual symmetries
\eqref{eq:Lie_pt} are now standard diffeomorphisms in jet space,
$z^{A}\mapsto\tilde{z}^{A}=\tilde{z}^{A}(z^{B};\varepsilon)$, having as
infinitesimal transformations
\begin{equation}\label{eq:Lie_ptjs}
\tilde{z}^{A}=z^{A}+\varepsilon{X^{A}}(z^{B})+\cdots,
\end{equation}
whose generator is a standard vector field in \emph{jet space}
\begin{equation}\label{eq:genjs}
\bm{X}=X^{A}(z^{B})\bm{\partial}_{A},\qquad
{X^{A}(z^{B})\equiv\frac{\partial\tilde{z}^{A}}{\partial\varepsilon}
\hspace{0.05in}\vline}_{\hspace{0.05in}\varepsilon=0}.
\end{equation}
The spacetime metric \eqref{eq:Ansatz} can be considered as an object in
jet space,
\begin{equation}\label{eq:genmet}
\bm{\mathrm{d}s^2}=g_{AB}(z^C)\bm{\mathrm{d}z}^A\bm{\mathrm{d}z}^B
=g_{\alpha\beta}(x^{\mu},u^J)
\bm{\mathrm{d}x}^{\alpha}\bm{\mathrm{d}x}^{\beta},
\end{equation}
with the trivial choices $g_{\mu{I}}=0=g_{IJ}$ for all $\mu$, and $I,J$.
Since the transformations are now standard one-parameter diffeomorphisms of
jet space, the residual symmetries are only standard isometries of the
jet space metric \eqref{eq:genmet}, i.e.\ the obvious criterion to
find residual symmetries in this geometrical approach is that the generators
are just jet space Killing fields,
\begin{equation}\label{eq:liederiv}
\bm{\pounds_{\!X} g}_{AB}
=X^{C}\partial_{C}g_{AB}+2g_{C(A}\partial_{B)}X^{C}=0.
\end{equation}
Using the fact that the jet space metric \eqref{eq:genmet} has only
nontrivial components along spacetime directions we exactly recover the
conditions \eqref{eq:critinf} already obtained by an elementary infinitesimal
reasoning.

The necessity of conditions \eqref{eq:critinf} is beyond doubt, indeed, it is
reinforced by the fact that the previous alternative geometrical viewpoint
gives exactly the same result. However, we are obliged to ask whether they
are sufficient to characterize the residual symmetries. Fortunately, the
geometrical picture brings some light on how to answer this question. What we
have learned so far is that residual symmetries are isometries of the trivial
lifting of the metric to jet space. It is very common in many
\emph{Ans{\"a}tze} that some of their structural functions do not depend on all
coordinates. Hence, there is a subset of the involved jet space
isometries for which the following anomalous scenario is possible: we can
compensate some spacetime diffeomorphisms with redefinitions which keep
invariant the form of the metric only formally, because they fail to respect
the original dependencies of the structural functions on spacetime
coordinates. Consequently, these are not residual symmetries in a strict
sense. In order to illustrate this situation we use again the Collinson
\emph{Ansatz} \eqref{eq:coll}, where all structural functions are independent of
$\tau$ and $\sigma$. Now, consider generalizing the rotation \eqref{eq:diff}
in the $(\tau,\sigma)$-plane to any other one-parameter reparameterization on
the same plane acting exclusively in the time coordinate
\begin{equation}
(\tau,\sigma,z)\,\mapsto\,
(\tilde{\tau}=\tilde{\tau}(\tau,\sigma;\varepsilon),
\tilde{\sigma}=\sigma, \tilde{z}=z).
\end{equation}
If one compensates for the previous diffeomorphism with the following
redefinitions,
\begin{align}
(a,b,P,Q)\,\mapsto\biggl(
\,\tilde{a}&=a\,\partial_\tau\tilde{\tau}-\partial_\sigma\tilde{\tau},
&\tilde{b}&=b\,\partial_\tau\tilde{\tau}+\partial_\sigma\tilde{\tau},
\nonumber\\
\tilde{P}&=P-\frac{1}{2}\ln\partial_\tau\tilde{\tau},
&\tilde{Q}&=Q+\frac{1}{2}\ln\partial_\tau\tilde{\tau}\biggr),
\label{eq:transcol}
\end{align}
it is easy to verify that the form of the metric remains unchanged as in
\eqref{eq:collb}, but the present transformed line element does not
represents anymore a manifestly stationary axisymmetric \emph{Ansatz} since the new
structural functions depend now on the new coordinates $\tilde{\tau}$ and
$\tilde{\sigma}$. However, the generator associated to the previous
transformation, which according to definitions \eqref{eq:compts} is
\begin{align}
\hspace{-0.07in}
\bm{X}={}&\xi^{\tau}(\tau,\sigma)\bm{\partial_{\tau}}
+\left(a\partial_{\tau}\xi^{\tau}
-\partial_{\sigma}\xi^{\tau}\right)\bm{\partial_{a}}
\nonumber\\
&+\left(b\partial_{\tau}\xi^{\tau}
+\partial_{\sigma}\xi^{\tau}\right)\bm{\partial_{b}}
-\frac{1}{2}\partial_{\tau}\xi^{\tau}\bm{\partial_{P}}
+\frac{1}{2}\partial_{\tau}\xi^{\tau}\bm{\partial_{Q}},
\label{eq:gencounterexa}
\end{align}
perfectly satisfies the vanishing Lie-derivative condition
\eqref{eq:liederiv}, or its dimensional reduction \eqref{eq:critinf},
evaluated on the Collinson \emph{Ansatz} \eqref{eq:coll}. Unfortunately, this is not
the only example of formal form-invariance; an exclusive reparameterization
of the angular coordinate $\sigma$ leads to similar results. The geometrical
approach is elegant and concise, but it is blind to the above problem if one
does not provide additional information. In summary, the conditions
\eqref{eq:critinf} are not sufficient to characterize strict form invariance.

In order to prevent situations as those described above, we need to supplement
the Lie-derivative conditions with others including the relevant information
on concrete dependencies. This is done by knowing how Lie-point
transformations \eqref{eq:Lie_pt} are prolonged to act on the derivatives
$u^I_\alpha\equiv{\partial{u}^I}/{\partial{x}^\alpha}$. In particular at the
infinitesimal level \eqref{eq:trans1} we have
\begin{equation}\label{eq:devinfini}
\tilde{u}^I_\alpha\equiv\frac{\partial\tilde{u}^I}{\partial\tilde{x}^\alpha}
=\frac{\mathrm{d}u^{I}+{\varepsilon}\mathrm{d}\eta^{I}+\cdots}
{\mathrm{d}x^\alpha+{\varepsilon}\mathrm{d}\xi^\alpha+\cdots}
=u^I_\alpha+\varepsilon\eta^I_\alpha+\cdots,
\end{equation}
where the prolongations are given in terms of the components of the original
generator as \cite{StH}
\begin{align}
\eta^I_\alpha\equiv{}&
{\frac{\partial\tilde{u}^I_\alpha}{\partial\varepsilon}
\hspace{0.05in}\vline}_{\hspace{0.05in}\varepsilon=0}\nonumber\\
={}&
\partial_\alpha\eta^I+u^J_\alpha\partial_J\eta^I
-u^I_\beta\left(\partial_\alpha\xi^\beta
+u^J_\alpha\partial_J\xi^\beta\right).\frac{}{}
\label{eq:prolcompts}
\end{align}
It is easy to see from here that even when the derivative of a given
structural function, let us say $u^{\bar{I}}$, with respect to a particular
coordinate, e.g.\ $x^{\hat{\alpha}}$, vanishes, their transformed counterpart
is not necessarily zero since it receives many other contributions. Suppose
we classify the coordinates and structural functions according to the
following: $x^{\alpha}=\left(x^{\bar{\alpha}}, x^{\hat{\alpha}}\right)$ and
$u^{I}=\left(u^{\bar{I}}, u^{\hat{I}}\right)$, where $u^{\bar{I}}$ denotes a
subset of structural functions which depend exclusively on the subset of
coordinates $x^{\bar{\alpha}}$, i.e.\
$u^{\bar{I}}=u^{\bar{I}}(x^{\bar{\alpha}})$. The rest of the functions and
coordinates are denoted by $u^{\hat{I}}$ and $x^{\hat{\alpha}}$,
respectively, where in general $u^{\hat{I}}=u^{\hat{I}}(x^{\bar{\alpha}},
x^{\hat{\alpha}})$. Then, if for the starting \emph{Ansatz}
$u^{\bar{I}}_{\hat{\alpha}}=0$, in order to avoid the described problem we
need to keep the same condition after the transformation, i.e.\
$\tilde{u}^{\bar{I}}_{\hat{\alpha}}=0$, which according to
\eqref{eq:devinfini} is infinitesimally equivalent to demanding
\begin{equation}
\eta^{\bar{I}}_{\hat{\alpha}}=\partial_{\hat{\alpha}}\eta^{\bar{I}}
+u^{\hat{J}}_{\hat{\alpha}}\partial_{\hat{J}}\eta^{\bar{I}}
-u^{\bar{I}}_{\bar{\beta}}\partial_{\hat{\alpha}}\xi^{\bar{\beta}}=0,
\end{equation}
where we have made use of conditions \eqref{eq:crit1}. Since this condition
must be satisfied for any values of the independent derivatives
$u^{\hat{J}}_{\hat{\alpha}}$ and $u^{\bar{I}}_{\bar{\beta}}$, their
companion coefficients must vanish independently, which defines the
definitive supplemental conditions
\begin{equation}
\partial_{\hat{\alpha}}\eta^{\bar{I}}
=\partial_{\hat{J}}\eta^{\bar{I}}
=\partial_{\hat{\alpha}}\xi^{\bar{\beta}}=0.
\end{equation}
These conditions establish that if a structural function $u^{\bar{I}}$ is
independent of some coordinates $x^{\hat{\alpha}}$, then their corresponding
generator component $\eta^{\bar{I}}$ is also independent of such coordinates
as well as of those structural functions that do depend on
$x^{\hat{\alpha}}$, and additionally that the components $\xi^{\bar{\beta}}$
along the other coordinates also are independent of $x^{\hat{\alpha}}$.

Summarizing, the full criterion defining the residual symmetries of a
gravitational \emph{Ansatz} is given by the next set of equations
\begin{subequations}\label{eq:crit}
\begin{align}
\xi^{\alpha}\partial_{\alpha}g_{\mu\nu}
+2g_{\alpha(\mu}\partial_{\nu)}\xi^{\alpha}
+\eta^{I}\partial_{I}g_{\mu\nu}
&=0,\frac{}{}\label{eq:crita}\\
\partial_{I}\xi^{\alpha}&=0,\label{eq:critb}\\
\partial_{\hat{\alpha}}\eta^{\bar{I}}
=\partial_{\hat{J}}\eta^{\bar{I}}
=\partial_{\hat{\alpha}}\xi^{\bar{\beta}}&=0.\label{eq:critxi}
\end{align}
\end{subequations}
As we will show in detail in the following section and in the appendixes
this criterion provides an effective computational procedure for finding all
the residual symmetries of almost any gravitational \emph{Ansatz} of interest. In
solving this linear system we will find integration constants or functions
that parametrize the components of the generator \eqref{eq:gengen}. It is an
easy task to show that the commutator of two residual symmetries generators
also satisfies the above conditions. Hence, the constants or functions
appearing in the integration process just span finite or infinite dimensional
Lie algebras. In other words the solution of the proposed system will be a
linear superposition of independent infinitesimal generators of residual
symmetries of the metric. As a byproduct, we consequently find the Killing
vectors of the corresponding metric as part of the solution; it is done in an
easier way than solving the standard Killing equations, where the \emph{Ansatz}
function are considered as dependent variables. Once we know such independent
infinitesimal generators we promote them to their corresponding finite
one-parameter transformations. For integrating the generators we use the
method of differential invariants described in appendix \ref{app:inv}. The
composition of all the one-parameter transformations gives rise to the most
general connected residual symmetry group for any gravitational \emph{Ansatz}. In
the following sections we examine the consequences of applying the proposed
criterion to several metric \emph{Ans\"{a}tze}, finding their corresponding residual
symmetry algebras and groups. We start with the illustrative example used so
far: the Collinson \emph{Ansatz}.

\section{\label{sec:Coll}The Collinson \emph{Ansatz}}

We now reconsider the Collinson \emph{Ansatz}, whose line element was given in the
introduction. The \emph{Ansatz} \eqref{eq:coll} was relevant at the beginning of the
search of interior solutions for the Kerr exterior spacetime. The so-called
Collinson theorem establishes that if a stationary axisymmetric spacetime is
also conformally flat then it must be necessarily static
\cite{Collinson:1976,Garcia:2002gj,AyonBeato:2005yx}. Hence, there are no
conformally flat Kerr interiors as opposed to the conformally flat interior
of the Schwarzschild metric. The virtue of this \emph{Ansatz} is that allows us to
write in a simple way the components of the Weyl tensor, producing an almost
straightforward integration of the conformally flat conditions.

The jet space parametrization of the Collinson \emph{Ansatz},
$z^{A}=\left(x^{\mu},u^{I}\right)$, consists of the spacetimes coordinates
$x^{\mu}=\left(\tau,\sigma,z,\bar{z}\right)$ together with the structural
functions $u^{I}=\left(a,b,P,Q\right)$. Moreover, the structural functions
are all independent of the coordinates
$x^{\hat{\alpha}}=\left(\tau,\sigma\right)$ and using the classification of
the previous section $u^{\bar{I}}=u^{I}$ ($u^{\hat{I}}=\{\emptyset\}$), i.e.\
all of them only depend on the coordinates
$x^{\bar{\alpha}}=\left(z,\bar{z}\right)$. This implies the complementary
residual conditions \eqref{eq:critxi} become
\begin{equation}\label{eq:colc}
\partial_{\tau}\eta^{I}=\partial_{\sigma}\eta^{I}
=\partial_{\tau}\xi^{z}=\partial_{\sigma}\xi^{z}=0.
\end{equation}
Using also the condition \eqref{eq:critb}, the general form of the
infinitesimal generator of residual symmetries for the Collinson \emph{Ansatz} is
the following
\begin{align}
\bm{X}={}&\xi^{\tau}(\tau,\sigma,z,\bar{z})\bm{\partial_\tau}
      +\xi^{\sigma}(\tau, \sigma,z,\bar{z})\bm{\partial_\sigma}
      +\xi^{z}(z,\bar{z})\bm{\partial_z}\nonumber\\
&+\overline{\xi^{z}}(z,\bar{z})\bm{\partial_{\bar{z}}}
 +\eta^{a}(z,\bar{z},a,b,P,Q)\bm{\partial_a}\nonumber\\
&+\eta^{b}(z,\bar{z},a,b,P,Q)\bm{\partial_b}
 +\eta^{P}(z,\bar{z},a,b,P,Q)\bm{\partial_P}\nonumber\\
&+\eta^{Q}(z,\bar{z},a,b,P,Q)\bm{\partial_Q},
\label{eq:gencol1}
\end{align}
where all the components are real except $\xi^{z}$, and $\overline{\xi^{z}}$
denotes its complex conjugate. We are now in position to implement the
remaining residual conditions (\ref{eq:crita}), which give the following
system of partial differential equations for the components of the previous
generator
\begin{subequations}\label{eq:resysColl}
\begin{align}
2\eta^{Q}+\frac{\left(\eta^{a}+\eta^{b}\right)}{a+b}
-2\partial_{\tau}\xi^{\tau}+(b-a)\partial_{\tau}\xi^{\sigma}&=0,
\label{eq:eq1} \\
(b-a)\eta^{Q}+\frac{\left( b\eta^{a}-a\eta^{b}\right)}{a+b}
+\partial_{\sigma}\xi^{\tau}-ab\partial_{\tau}\xi^{\sigma}&
\nonumber\\
-\frac{1}{2}\left( b-a \right)\left(\partial_{\tau}\xi^{\tau}
+\partial_{\sigma}\xi^{\sigma} \right)&=0,\label{eq:eq2} \\
{}2ab\eta^{Q}-\frac{\left(b^2\eta^{a}+a^2\eta^{b}\right)}{a+b}
-(b-a)\partial_{\sigma}\xi^{\tau}& \nonumber\\
{}-2ab\partial_{\sigma}\xi^{\sigma}&=0,
\frac{}{}\label{eq:eq3} \\
\eta^{Q}+\eta^{P}-\frac{1}{2}\left( \partial_{z}\xi^{z}
+\partial_{\bar{z}}\bar{\xi^{z}} \right)&=0,\label{eq:eq4} \\
(b-a)\partial_{z}\xi^{\sigma}-2\partial_{z}\xi^{\tau}&=0,
\frac{}{}\label{eq:eq6} \\
(b-a)\partial_{z}\xi^{\tau}+2ab\partial_{z}\xi^{\sigma}&=0,
\frac{}{}\label{eq:eq7} \\
\partial_{\bar{z}}\xi^{z}&=0,\frac{}{}\label{eq:eq9}
\end{align}
\end{subequations}
where the complex conjugate of
Eqs.~(\ref{eq:eq6})-(\ref{eq:eq9}) must also be included.

In order to integrate the above system we need to remember that in this
context the structural functions of the \emph{Ansatz} and the spacetime coordinates
are all assumed as independent variables. This turns the integration process
into a very easy job; e.g.\ in Eqs.~(\ref{eq:eq6}) and (\ref{eq:eq7}), since
the spacetime components of the generator are independent of the structural
functions, the only way in which these equations hold is if the coefficients
in front of all the functionally independent expressions of the structural
functions vanish separately
\begin{equation}\label{eq:xisapp}
\partial_{z}\xi^{\tau}=0=\partial_{z}\xi^{\sigma}.
\end{equation}
Since the complex conjugates of these conditions are also valid we conclude
that $\xi^{\tau}=\xi^{\tau}(\tau,\sigma)$ and
$\xi^{\sigma}=\xi^{\sigma}(\tau,\sigma)$. Additionally, equation
\eqref{eq:eq9} implies the component along the complex plane is holomorphic
$\xi^{z}=\xi^{z}(z)$. Another consequence of the independency between the
structural functions and the coordinates can be obtained when
Eqs.~\eqref{eq:eq1}-\eqref{eq:eq4} are derived with respect to the Killing
coordinates $\tau$ and $\sigma$ and the conditions \eqref{eq:colc} are used;
the resulting equations are only satisfied if the following derivatives are
constant
\begin{align}
\partial_{\tau}\xi^{\tau}&=\hat{C}_{1},   &
\partial_{\sigma}\xi^{\tau}&=C_{3},\nonumber\\
\partial_{\sigma}\xi^{\sigma}&=\hat{C}_{2}, &
\partial_{\tau}\xi^{\sigma}&=C_{4}.
\label{eq:sysxitxis}
\end{align}
This system has the trivial solution
\begin{subequations}
\begin{align}
\xi^{\tau}&=\hat{C}_{1}\tau+C_{3}\sigma+\kappa_{1},\label{eq:xitcol} \\
\xi^{\sigma}&=\hat{C}_{2}\sigma+C_{4}\tau+\kappa_{2}.\label{eq:xiscol}
\end{align}
\end{subequations}
If one replaces now the set of conditions \eqref{eq:sysxitxis} or their
solutions in Eqs.~\eqref{eq:eq1}-\eqref{eq:eq4}, a linear algebraic system of
equations for the $\eta^{I}$ is obtained whose solution is given below
\begin{subequations}
\begin{align}
\eta^{a}&=C_{4}a^{2}+C_{2}a-C_{3},\label{eq:etaacol} \\
\eta^{b}&=-C_{4}b^{2}+C_{2}b+C_{3},\label{eq:etabcol} \\
\eta^{P}&=\frac{1}{2}(\partial_{z}\xi^{z}
+\partial_{\bar{z}}\bar{\xi^{z}}-C_{1}),\label{eq:etapcol}\\
\eta^{Q}&=\frac{1}{2}C_{1},\label{eq:etaqcol}
\end{align}
\end{subequations}
where we have redefined two of the integration constants as
$C_{1}=\hat{C}_1+\hat{C}_2$ and $C_{2}=\hat{C}_1-\hat{C}_2$. This ends the
integration of the residual system \eqref{eq:resysColl}, giving the following
generator for the residual symmetries of the Collinson \emph{Ansatz}
\begin{align}
\bm{X}={}&C_{1}(\tau\bm{\partial_{\tau}}
+\sigma\bm{\partial_{\sigma}}+\bm{\partial_{Q}}
-\bm{\partial_{P}})\nonumber\\
&+C_{2}(\tau\bm{\partial_{\tau}}
-\sigma\bm{\partial_{\sigma}}+2a\bm{\partial_{a}}
+2b\bm{\partial_{b}})\nonumber\\
&+C_{3}(\sigma\bm{\partial_{\tau}}-\bm{\partial_{a}}
+\bm{\partial_{b}})\nonumber\\
&+C_{4}(\tau\bm{\partial_{\sigma}}+a^{2}\bm{\partial_{a}}
-b^{2}\bm{\partial_{b}})\nonumber\\
&+\xi^{z}\bm{\partial_{z}}
+\overline{\xi^{z}}\bm{\partial_{\bar{z}}}
+\frac{1}{2}(\partial_{z}\xi^{z}
+\partial_{\bar{z}}\overline{\xi^{z}})\bm{\partial_{P}}
\nonumber\\
&+\kappa_{1}\bm{\partial_{\tau}}
+\kappa_{2}\bm{\partial_{\sigma}}.
\label{eq:gencol2}
\end{align}

As we anticipate in the previous section the resulting solution is a linear
combination of vector fields, each one generating a one-parameter group of
residual symmetries. We label this individual generators as follows
\begin{subequations}\label{eq:colgens}
\begin{align}
\bm{X}_{1}&=\tau\bm{\partial_{\tau}}+\sigma\bm{\partial_{\sigma}}
-\bm{\partial_{P}}+\bm{\partial_{Q}},
\label{eq:x1col}\\
\bm{X}_{2}&=\tau\bm{\partial_{\tau}}-\sigma\bm{\partial_{\sigma}}
+2a\bm{\partial_{a}}+2b\bm{\partial_{b}},
\label{eq:x2col}\\
\bm{X}_{3}&=\sigma\bm{\partial_{\tau}}-\bm{\partial_{a}}
+\bm{\partial_{b}},
\label{eq:x3col}\\
\bm{X}_{4}&=\tau\bm{\partial_{\sigma}}+a^{2}\bm{\partial_{a}}
-b^{2}\bm{\partial_{b}},
\label{eq:x4col}\\
\bm{\hat{X}}_{\xi^{z}}&=\xi^{z}\bm{\partial_{z}}
+\overline{\xi^{z}}\bm{\partial_{\bar{z}}}
+\frac{1}{2}\left(\partial_{z}\xi^{z}
+\partial_{\bar{z}}\overline{\xi^{z}}\right)\bm{\partial_{P}},
\label{eq:xxicol}\\
\bm{k}&=\bm{\partial_{\tau}},\label{eq:kill1col}\\
\bm{m}&=\bm{\partial_{\sigma}},\label{eq:kill2col}
\end{align}
\end{subequations}
where as was pointed out previously, $\xi^z=\xi^z(z)$ is an arbitrary
holomorphic function and $\overline{\xi^{z}}$ is its complex conjugate. The
generators \eqref{eq:x1col}-\eqref{eq:xxicol} are associated to the residual
symmetries of the Collinson \emph{Ansatz} while \eqref{eq:kill1col} and
\eqref{eq:kill2col} are their corresponding Killing vectors. On the one hand,
these Killing vectors together with the first four generators
\eqref{eq:x1col}-\eqref{eq:x4col} form a six-dimensional Lie subalgebra. On
the other hand, the generator \eqref{eq:xxicol} represents an infinite
dimensional Lie subalgebra. The concrete Lie algebra is characterized by
TABLE~\ref{table:Coll}, where the entry for a given row and column represents
the corresponding commutator.%
\begin{table}[hbtp!]
\caption{\label{table:Coll} The commutator table for a Collinson
\emph{Ansatz}.}
\begin{ruledtabular}
\begin{tabular}{c|ccccccc}
& $\bm{X}_1$ & $\bm{X}_2$ & $\bm{X}_3$ & $\bm{X}_4$ & $\bm{k}$ & $\bm{m}$
& $\bm{\hat{X}}_{\xi^z_2}$ \\
\noalign{\smallskip}\hline\noalign{\smallskip}
$\bm{X}_1$ & 0 & 0 & 0 & 0 & $-\bm{k}$ & $-\bm{m}$ & 0 \\
$\bm{X}_2$ & 0 & 0 & $-2\bm{X}_3$ & $2\bm{X}_4$ & $-\bm{k}$ & $\bm{m}$ & 0 \\
$\bm{X}_3$ & 0 & $2\bm{X}_3$ & 0 & $-\bm{X}_2$ & 0 & $-\bm{k}$ & 0 \\
$\bm{X}_4$ & 0 & $-2\bm{X}_4$ & $\bm{X}_2$ & 0 & $-\bm{m}$ & 0 & 0 \\
$\bm{k}$   & $\bm{k}$ & $\bm{k}$ & 0 & $\bm{m}$ & 0 & 0 & 0 \\
$\bm{m}$   & $\bm{m}$ & $-\bm{m}$ & $\bm{k}$ & 0 & 0 & 0 & 0 \\
$\bm{\hat{X}}_{\xi^z_1}$
           & 0 & 0 & 0 & 0 & 0 & 0 & $\bm{\hat{X}}_{\Omega^{z}}$ \\
\end{tabular}
\end{ruledtabular}
\end{table}

The infinite-dimensional subalgebra can be characterized as follows, in
TABLE~\ref{table:Coll} we denote
\begin{equation}
\Omega^{z}=\xi^{z}_{1}\partial_{z}\xi^{z}_2-\xi^{z}_{2}\partial_{z}\xi^{z}_1.
\label{eq:Omega}
\end{equation}
Since $\xi^{z}$ is a holomorphic function we can expand it by a Laurent
series,
\begin{align}
\xi^{z}(z)&=\sum_{n=-\infty}^{\infty}a_{n}(-z^{n+1}),
\label{eq:Ls}
\end{align}
and the generator \eqref{eq:xxicol} becomes an infinite expansion
\begin{equation}
\bm{\hat{X}}_{\xi^{z}}=\sum_{n=-\infty}^{\infty}[a_{n}\bm{L}_{n}
+\bar{a}_{n}\bm{\bar{L}}_{n}],
\end{equation}
on the complex generators
\begin{align}
\bm{L}_{n}&=-z^{n+1}\bm{\partial}_{z}-\frac{n+1}{2}z^{n}\bm{\partial}_{P},
\label{eq:ln}
\end{align}
and their complex conjugate $\bm{\bar{L}}_{n}$. Both sets of vector fields
obey the following algebra
\begin{align}
\left[\bm{L}_{n},\bm{L}_{m}\right]&=(n-m)\bm{L}_{n+m},\label{eq:alcon1}\\
\left[\bm{\bar{L}}_{n},\bm{\bar{L}}_{m}\right]&=(n-m)\bm{\bar{L}}_{n+m},
\label{eq:alcon2}
\end{align}
i.e.\ the infinite-dimensional subalgebra of the residual symmetries of the
Collinson \emph{Ansatz} is composed of just two copies of the Witt algebra
\cite{ScM}, i.e.\ the so-called conformal algebra without the central
extension.

Now we proceed to integrate the various infinitesimal transformations
generated by the vector fields \eqref{eq:colgens}, in order to find their
corresponding one-parameter finite transformations. We made use of the method
of differential invariants explained in appendix \ref{app:inv}, where it is
explicitly show how to integrate several generators which repeatedly appears
in most of the examples. Hence, we do not include the specific details of
each case here. The case of generator \eqref{eq:xxicol} is slightly different
since it involves the use of an arbitrary function of the $z$ coordinate, but
their integration is also included in appendix \ref{app:inv} as an example of
how to deal with these kind of generators. Using the method of this appendix,
the finite transformation found by integrating the generator (\ref{eq:x1col})
is the following
\begin{align}
\tilde{\tau}&=\lambda\tau, & \tilde{\sigma}&=\lambda\sigma, &
\tilde{z}&=z, \nonumber\\
\tilde{a}&=a,              & \tilde{b}&=b, \nonumber\\
\tilde{P}&=P-\ln(\lambda), & \tilde{Q}&=Q+\ln(\lambda),
\label{eq:t1col}
\end{align}
where $\lambda=\exp{\varepsilon}$. In other words, if the Killing coordinates
$\tau$ and $\sigma$ scale in the same way, this is compensated for by appropriate
translations along the functions $P$ and $Q$. For the generator
(\ref{eq:x2col}) we get the transformation
\begin{align}
\tilde{\tau}&=\lambda\tau, &
\tilde{\sigma}&=\lambda^{-1}\sigma, &
\tilde{z}&=z, \nonumber\\
\tilde{a}&=\lambda^{2}a,   & \tilde{b}&=\lambda^{2}b, &
\tilde{P}&=P, & \tilde{Q}&=Q.
\label{eq:t2col}
\end{align}
This time we see that if $\tau$ and $\sigma$ scale inversely, this must be
compensated for with a double scaling in the functions $a$ and $b$. The
transformation arising from the exponentiation of the vector field
(\ref{eq:x3col}) is
\begin{align}
\tilde{\tau}&=\tau+\varepsilon\sigma, & \tilde{\sigma}&=\sigma, &
\tilde{z}&=z,\nonumber\\
\tilde{a}&=a-\varepsilon, & \tilde{b}&=b+\varepsilon, &
\tilde{P}&=P, & \tilde{Q}&=Q,
\label{eq:t3col}
\end{align}
which is precisely the residual symmetry we use as example at the beginning.
As was already emphasized, this transformation establishes that the effects
of rotating the time coordinate $\tau$ in the Killing plane $(\tau,\sigma)$
can be eliminated by translations with different sign in the functions $a$
and $b$. Now, we present the residual symmetry that we get after integrating
the generator \eqref{eq:x4col}
\begin{align}
\tilde{\tau}&=\tau, & \tilde{\sigma}&=\sigma+\varepsilon\tau, &
\tilde{z}&=z, \nonumber\\
\tilde{a}&=\frac{a}{1-\varepsilon a}, &
\tilde{b}&=\frac{b}{1+\varepsilon b}, &
\tilde{P}&=P, & \tilde{Q}&=Q,\label{eq:t4col}
\end{align}
i.e.\ if one instead rotates the angle $\sigma$ in the Killing plane it is
neutralized by appropriate special conformal transformations on the functions
$a$ and $b$. Finally, the case of the generator (\ref{eq:xxicol}) is studied
in detail in appendix \ref{app:inv}, where it is obtained the following
finite transformation
\begin{align}
\tilde{\tau}&=\tau, & \tilde{\sigma}&=\sigma, &
\tilde{z}&=\tilde{z}(z),\nonumber\\
\tilde{a}&=a, & \tilde{b}&=b, &
\tilde{P}&=P+\ln\left|\frac{\mathrm{d}\tilde{z}}
                           {\mathrm{d}z}\right|, &
\tilde{Q}&=Q. \label{eq:t5col}
\end{align}
In this case, any conformal transformation of the complex $z$ plane will be
compensated for with a supertranslation of the conformal function $P$, involving
the nonvanishing derivative of the holomorphic transformation in $z$.

We are ready to write now the most general connected residual transformation
that the Collinson \emph{Ansatz} (\ref{eq:coll}) admits. It is defined by the
composition of the transformations (\ref{eq:t1col})-(\ref{eq:t5col}) and is
given by
\begin{align}
\tilde{\tau}&=\alpha\tau+\beta\sigma, \quad
\tilde{\sigma}=\delta\sigma+\gamma\tau, &
\tilde{z}&=\tilde{z}(z), \nonumber\\
\tilde{a}&=\frac{\alpha{a}-\beta}
                {-\gamma{a}+\delta}, &
\tilde{b}&=\frac{\alpha{b}+\beta}
                {\gamma{b}+\delta},\nonumber\\
\tilde{P}&=P+\ln\left|\frac1{\sqrt{\alpha\delta-\beta\gamma}}
\frac{\mathrm{d}\tilde{z}}{\mathrm{d}z}\right|, &
\tilde{Q}&=Q+\ln\sqrt{\alpha\delta-\beta\gamma},\label{eq:tgencol}
\end{align}
where $\alpha\delta-\beta\gamma>0$. This transformation tell us that the Collinson \emph{Ansatz} is form invariant under
two-dimensional general linear transformations, with positive determinant, on the Killing
$(\tau,\sigma)$ plane and arbitrary conformal transformations on the complex
$z$ plane, provided that the structural functions will be redefined according
to Eqs.~(\ref{eq:tgencol}). These are precisely the transformations actively
used in Refs.~\cite{Collinson:1976,Garcia:2002gj,AyonBeato:2005yx}.

In the following sections we study many other spacetime examples emphasizing
just the main results regarding their residual symmetries, the corresponding
details can be found in the appendixes.

\section{\label{sec:Spher}Spherically symmetric \emph{Ansatz}}

We follow up now by studying the very well-known and studied case of spherically
symmetric spacetimes. In their static version they are described by the
following metric
\begin{align}
\bm{\mathrm{d}s^{2}}={}&-N^{2}(r)F(r)\bm{\mathrm{d}t^{2}}
+\frac{\bm{\mathrm{d}r^{2}}}{F(r)}\nonumber\\
&+Y^{2}(r)(\bm{\mathrm{d}\theta^{2}}+\sin^{2}\theta\,
\bm{\mathrm{d}\varphi^{2}}),
\label{eq:ssa}
\end{align}
where a gauge election remains to be done. Since such election is done in
practice in many different ways, we prefer to keep this freedom in order that
our analysis can be adapted to the different choices. The involved jet
space coordinates $z^{A}=\left(x^{\mu},u^{I}\right)$ are now the spacetime
spherical coordinates $x^{\mu}=\left(t,r,\theta,\varphi\right)$ and the
metric functions $u^{I}=\left(N,F,Y\right)$. The infinitesimal generators of
the corresponding residual symmetries are studied in detail in
App.~\ref{app:Spher}, the final result is
\begin{subequations}\label{eq:gentssa}
\begin{align}
\bm{X}_{1}&=t\bm{\partial_{t}}-N\bm{\partial_{N}},\label{eq:x1ssa}\\
\bm{\hat{X}}_{\xi^{r}}&=\xi^{r}(r)\bm{\partial_{r}}
+\partial_{r}\xi^{r}\left(2F\bm{\partial_{F}}
-N\bm{\partial_{N}}\right),\label{eq:xrssa}\\
\bm{L}_{x}&=\sin\varphi\bm{\partial_{\theta}}
+\cot\theta \cos\varphi\bm{\partial_{\varphi}},\label{eq:kil1ssa}\\
\bm{L}_{y}&=\cos\varphi\bm{\partial_{\theta}}
-\cot\theta \sin\varphi\bm{\partial_{\varphi}},\label{eq:kil2ssa}\\
\bm{L}_{z}&=\bm{\partial_{\varphi}},\label{eq:kil3ssa}\\
\bm{k}&=\bm{\partial_{t}}.\label{eq:kil4ssa}
\end{align}
\end{subequations}
The generators \eqref{eq:kil1ssa}-\eqref{eq:kil4ssa} are just the spherically
symmetric and stationary Killing vectors of the metric, therefore there are
only two generators strictly corresponding to residual symmetries and one of
them spans an infinite-dimensional subalgebra.%
\begin{table}[hbtp!]
\caption{\label{table:sssa} The commutator table for a static spherical \emph{Ansatz}.}
\begin{ruledtabular}
\begin{tabular}{c|cccccc}
& $\bm{X}_1$ & $\bm{k}$ & $\bm{L}_x$ & $\bm{L}_y$ & $\bm{L}_z$ &
$\bm{\hat{X}}_{\xi^r_2}$ \\
\noalign{\smallskip}\hline\noalign{\smallskip}
$\bm{X}_1$ & 0 & $-\bm{k}$ & 0 & 0 & 0 & 0  \\
$\bm{k}$ & $\bm{k}$ & 0 & 0 & 0 & 0 & 0  \\
$\bm{L}_x$ & 0 & 0 & 0 & $\bm{L}_z$ & $-\bm{L}_y$ & 0  \\
$\bm{L}_y$ & 0 & 0 & $-\bm{L}_z$ & 0 & $\bm{L}_{x}$ & 0  \\
$\bm{L}_z$   & 0 & 0 & $\bm{L}_y$ & $-\bm{L}_x$ & 0 & 0  \\
$\bm{\hat{X}}_{\xi^r_1}$
           & 0 & 0 & 0 & 0 & 0 & $\bm{\hat{X}}_{\Omega^{r}}$ \\
\end{tabular}
\end{ruledtabular}
\end{table}
The generators \eqref{eq:gentssa} give rise to the algebra shown in
TABLE~\ref{table:sssa}.

The infinite-dimensional subalgebra is obtained from the Lie bracket of the
family of vector fields $\bm{\hat{X}}_{\xi^{r}}$ among themselves, where we
define
\begin{equation}\label{eq:Lambda}
\Omega^{r}=\xi^{r}_{1}\partial_{r}\xi^{r}_{2}
-\xi^{r}_{2}\partial_{r}\xi^{r}_{1}.
\end{equation}
Expanding the function $\xi^{r}$ as a Fourier series,
\begin{equation}\label{eq:Fsxir}
\xi^{r}(r)=\sum_{n=-\infty}^{\infty}a_{n}(-i\mathrm{e}^{-inr}),
\end{equation}
the generator \eqref{eq:xrssa} is written as a linear combination of the
infinite number of vector fields
\begin{equation}
\bm{L}_{n}=-i\mathrm{e}^{-inr}\left[\bm{\partial_{r}}
+in\left(N\bm{\partial_{N}}
-2F\bm{\partial_{F}}\right)\right],
\end{equation}
whose lie brackets obey the Witt algebra \eqref{eq:alcon1}.

Let us focus now in the finite transformations associated to the residual
symmetries that we find using the methods illustrated in App.~\ref{app:inv}.
The transformation found by integrating the generator (\ref{eq:x1ssa}) is
\begin{align}
\tilde{t}&=\lambda{t}, & \tilde{r}&=r, & \tilde{\theta}&=\theta, &
\tilde{\varphi}&=\varphi,\nonumber\\
\widetilde{F}&=F, & \widetilde{N}&=\lambda^{-1}N, & \widetilde{Y}&=Y,
\label{eq:t1ssa}
\end{align}
which tells us that any rescaling on time is compensated for with the inverse
rescaling on the function $N$; a very well-known and -used property of the
spherically symmetric \emph{Ansatz}. The family of generators (\ref{eq:xrssa})
involving a general dependence must be treated similarly to the last example
of App.~\ref{app:inv}. We get the finite transformations
\begin{align}
\tilde{t}=t,& &
\tilde{r}&=\tilde{r}(r), \quad\,\, \tilde{\theta}=\theta,
& \tilde{\varphi}&=\varphi,\nonumber\\
\widetilde{F}=\left(\frac{\mathrm{d}\tilde{r}}{\mathrm{d}r}\right)^{2}F,& &
\widetilde{N}&=\left(\frac{\mathrm{d}\tilde{r}}{\mathrm{d}{r}}\right)^{-1}N, &
\widetilde{Y}&=Y,\label{eq:t2ssa}
\end{align}
encoding that an arbitrary radial reparametrization is compensated for with a
local rescaling in the functions $N$ and $F$ involving the derivative of the
radial transformation. Consequently, the general residual transformation
admitted by the static spherical metric \emph{Ansatz} (\ref{eq:ssa}) is given by
\begin{align}
\tilde{t}=\lambda t,& & 
\tilde{r}&=\tilde{r}(r), \quad\,\, \tilde{\theta}=\theta, & 
\tilde{\varphi}&=\varphi,\nonumber\\
\widetilde{F}=\left(\frac{\mathrm{d}\tilde{r}}{\mathrm{d}r}\right)^{2}F,& &
\widetilde{N}&=\left(\lambda\frac{\mathrm{d}\tilde{r}}
{\mathrm{d}r}\right)^{-1}N, & 
\widetilde{Y}&=Y.\label{eq:tgenssa}
\end{align}

It is illustrative to analyze how the above results are generalized when the
spherically symmetric metric is allowed to depend on time
\begin{align}
\bm{\mathrm{d}s}^{2}={}&-N^{2}(t,r)F(r,t)\bm{\mathrm{d}t}^{2}
+\frac{\bm{\mathrm{d}r}^{2}}{F(t,r)}\nonumber\\
&+Y^{2}(t,r)\left(\bm{\mathrm{d}\theta}^{2}
+\sin^{2}\theta{\bm{\mathrm{d}\varphi}^{2}}\right).
\label{eq:ss}
\end{align}
The coordinates for the jet space are the same as the ones for the
static case, but as now the structural functions also depend on time the
supplementary conditions \eqref{eq:critxi} are different, see
App.~\ref{app:Spher}. The criterion gives the following two families of
generators for the residual symmetries
\begin{subequations}\label{eq:gentss}
\begin{align}
\bm{\check{X}}_{\xi^{t}}&=\xi^{t}(t)\bm{\partial_{t}}
-\partial_{t}\xi^{t}N\bm{\partial_{N}},\label{eq:xtss}\\
\bm{\hat{X}}_{\xi^{r}}&=\xi^{r}(r)\bm{\partial_{r}}
-\partial_{r}\xi^{r}\left(N\bm{\partial_{N}}-2F\bm{\partial_{F}}\right),
\label{eq:xrss}
\end{align}
\end{subequations}
together with the Killing vectors of the sphere
\eqref{eq:kil1ssa}-\eqref{eq:kil3ssa}. This means that loosing the stationary
symmetry \eqref{eq:kil4ssa} enhances the time scaling \eqref{eq:x1ssa} to a
time reparametrization. The Lie algebra spanned by the involved generators
is presented in TABLE~\ref{table:tssa}.%
\begin{table}[hbtp!]
\caption{\label{table:tssa} The commutator table for a spherical \emph{Ansatz}.}
\begin{ruledtabular}
\begin{tabular}{c|ccccc}
 & $\bm{L}_x$ & $\bm{L}_y$ & $\bm{L}_z$ & $\bm{\check{X}}_{\xi^{t}_2}$ &
 $\bm{\hat{X}}_{\xi^r_2}$ \\
\noalign{\smallskip}\hline\noalign{\smallskip}
$\bm{L}_x$  & 0 & $\bm{L}_z$ & $-\bm{L}_y$ & 0 & 0  \\
$\bm{L}_y$  & $-\bm{L}_z$ & 0 & $\bm{L}_{x}$ & 0 & 0  \\
$\bm{L}_z$  & $\bm{L}_y$ & $-\bm{L}_x$ & 0 & 0 & 0  \\
$\bm{\check{X}}_{\xi^{t}_{1}}$
            & 0 & 0 & 0 & $\bm{\check{X}}_{\Omega^{t}}$ & 0  \\
$\bm{\hat{X}}_{\xi^r_1}$ & 0 & 0 & 0 & 0 & $\bm{\hat{X}}_{\Omega^{r}}$ \\
\end{tabular}
\end{ruledtabular}
\end{table}
This time we have two commuting infinite-dimensional subalgebras, and the
quantities defining them are
\begin{equation}
\Omega^{k}=\xi^{k}_{1}\partial_{k}\xi^{k}_{2}
          -\xi^{k}_{2}\partial_{k}\xi^{k}_{1},
\label{eq:lambda2}
\end{equation}
where $k$ is a generic label for the coordinates $t$ or $r$, and no
summation over $k$ is understood. The generator \eqref{eq:xrss} is the same
appearing in the static case \eqref{eq:xrssa}. Hence, expanding the function
$\xi^{r}(r)$ in a Fourier series \eqref{eq:Fsxir} we obtain a Witt algebra. A
similar expansion for the function $\xi^{t}(t)$ of the other generator
\eqref{eq:xtss} gives a second copy of the Witt algebra satisfied by the
generators
\begin{equation}
\bm{\check{L}}_{n}=-i\mathrm{e}^{-int}\left(\bm{\partial_{r}}
+inN\bm{\partial_{N}}\right).
\end{equation}
These two copies are a manifestation of the conformal algebra without central
charge.

The finite symmetry arising from generator (\ref{eq:xtss}) is
\begin{align}
\tilde{t}&=\tilde{t}(t), & \tilde{r}&=r, \qquad \tilde{\theta}=\theta, &
\tilde{\varphi}&=\varphi,\nonumber\\
\tilde{F}&=F, &
\tilde{N}&=\left(\frac{\mathrm{d}\tilde{t}}{\mathrm{d}t}\right)^{-1}N, &
\tilde{Y}&=Y,\label{eq:t1ss}
\end{align}
i.e.\ any time reparametrization is compensated for with a local scaling of the
function $N$. The finite residual transformation that results from the
exponentiation of the vector field \eqref{eq:xrss} is exactly the one
displayed in the static case \eqref{eq:t2ssa}, with the obvious difference
that in the present case $F$ and $N$ depend also on time. The composition of
both transformations gives the most general connected residual symmetry
allowed by a spherically symmetric gravitational \emph{Ansatz}
\begin{align}
\tilde{t}=\tilde{t}(t),& &
\tilde{r}&=\tilde{r}(r), \quad\,\, \tilde{\theta}=\theta, &
\tilde{\varphi}&=\varphi,\nonumber\\
\widetilde{F}=\left(\frac{\mathrm{d}\tilde{r}}{\mathrm{d}r}\right)^{2}F,& &
\widetilde{N}&=\left(\frac{\mathrm{d}\tilde{t}}{\mathrm{d}t}
\frac{\mathrm{d}\tilde{r}}{\mathrm{d}r}\right)^{-1}N, &
\widetilde{Y}&=Y.\label{eq:trgenssa}
\end{align}
As indicated by the related infinite algebras these transformations are the
conformal freedom enjoyed by the Lorentzian fibers orthogonal to the spheres.

\section{\label{sec:AdS}AdS waves}

We study now the so-called AdS waves \cite{Siklos:1985,AyonBeato:2005qq}.
They have proved to be a very useful tool to inspect the dynamical and
asymptotic properties of highly nontrivial theories since they represent
exact realizations of the propagating degrees of freedoms of gravity in
several contexts
\cite{Siklos:1985,AyonBeato:2005qq,AyonBeato:2004fq,AyonBeato:2009yq}. An AdS
wave in $D$ dimensions is given by the metric
\begin{equation}\label{eq:AdSw}
\bm{\mathrm{d}s}^{2}=\frac{l^{2}}{y^{2}}\left[
-F(u,y,\vec{x})\bm{\mathrm{d}u}^{2}-2\bm{\mathrm{d}u\mathrm{d}v}
+\bm{\mathrm{d}y}^{2}+\bm{\mathrm{d}\bm{\vec{x}}}^{2}\right],
\end{equation}
where $\vec{x}$ is a $D-3$ Euclidean vector. The existence of residual
symmetries of this \emph{Ansatz} in lower dimension is already known
\cite{Siklos:1985,AyonBeato:2005qq}, and has been exploited to gauge away the
non-propagating degrees of freedoms of standard gravity in lower dimensions.
Nevertheless, the uniqueness of such symmetries has not been established.
Additionally, little is known about their general behavior in higher
dimensions, except that their famous three-dimensional relation with the
Virasoro algebra \cite{Brown:1986nw} can be extended to any dimension
\cite{Banados:1999tw}. It is our interest in this section to apply the
derived criterion to the AdS waves to fully investigate these questions. The
related jet space coordinates are $z^{A}=\left(x^{\mu},u^{I}\right)$
where $x^{\mu}=\left(u,v,y,\vec{x}\right)$, $u^{I}=(F)$. The infinitesimal
generators of residual symmetries are thoroughly studied in
App.~\ref{app:AdS}, and the final result is
\begin{subequations}\label{eq:gensads}
\begin{align}
\bm{X}_{1}={}&2v\bm{\partial_{v}}+y\bm{\partial_{y}}
+x^{i}\bm{\partial_{i}}+2F\bm{\partial_{F}},\label{eq:x1ads}\\
\bm{J_}{ij}={}&x_{i}\bm{\partial}_{j}-x_{j}\bm{\partial}_{i},
\label{eq:kil1ads}\\
\bm{\tilde{X}}_{\alpha}={}&\alpha(u)\bm{\partial_{v}}
-2\dot{\alpha}\bm{\partial_{F}},\label{eq:xaads}\\
\bm{\check{X}}_{\xi^{u}}={}&\xi^{u}(u)\bm{\partial_{u}}
+\frac{{\dot{\xi}^{u}}}{2}\left(y\bm{\partial_y}+x^{i}\bm{\partial_{i}}
-2F\bm{\partial_{F}}\right)\nonumber\\
&+\frac{1}{4}(y^2+\vec{x}^2)\bm{\tilde{X}}_{\ddot{\xi}^{u}},\label{eq:xuads}\\
\bm{X_}{\vec{P}}={}&\bm{\vec{P}}(u)+\vec{x}\cdot\!\bm{\tilde{X}}_{\dot{\vec{P}}},
\quad\bm{\vec{P}}(u)=P^{i}(u)\bm{\partial_{i}}.\label{eq:kil2ads}
\end{align}
\end{subequations}
The null Killing vector of the AdS waves, $\bm{\partial_{v}}$, corresponds to
the particular case of the generator \eqref{eq:xaads} when the function
$\alpha$ is a constant. It is important to emphasize that the rotations
\eqref{eq:kil1ads} are not isometries, since in general the structural
function $F$ in metric \eqref{eq:AdSw} is not rotation invariant, but will
change respecting the standard diffeomorphism rule and consequently leaving
the \emph{Ansatz} form invariant.%
\begin{table}[hbtp!]
\caption{\label{table:AdS} The commutator table for AdS waves.}
\begin{ruledtabular}
\begin{tabular}{c|ccccc}
&$\bm{X}_1$ & $\bm{J}_{kl}$ & $\bm{\tilde{X}}_{\beta}$ &
$\bm{\check{X}}_{\xi^{u}_{2}}$ & $\bm{X}_{\vec{Q}}$  \\
\noalign{\smallskip}\hline\noalign{\smallskip}
$\bm{X}_1$ & 0 & 0 & $-2\bm{\tilde{X}}_{\beta}$ & 0 & $-\bm{X}_{\vec{Q}}$ \\
$\bm{J}_{ij}$ & 0 & ${c_{ij,kl}}^{mn}\bm{J}_{mn}$ & 0 & 0 &
$-\bm{X}_{\vec{Q}_{ij}}$  \\
$\bm{\tilde{X}}_{\alpha}$ & $2\bm{\tilde{X}}_{\alpha}$ & 0 & 0 &
$-\bm{\tilde{X}}_{\dot{\alpha}\xi^{u}_2 }$ & 0 \\
$\bm{\check{X}}_{\xi^{u}_1}$   & 0 & 0 &
$\bm{\tilde{X}}_{\dot{\beta}\xi^{u}_{1}}$ &
$\bm{\check{X}}_{\Omega^{u}}$ &
$\bm{X}_{\vec{\gamma}_{1}\left[\vec{Q}\right]}$ \\
$\bm{X}_{\vec{P}}$ & $\bm{X}_{\vec{P}}$ & $\bm{X}_{\vec{P}_{kl}}$ & 0 &
$-\bm{X}_{\vec{\gamma}_{2}\left[\vec{P}\right]}$ &
$\bm{\tilde{X}}_{\vec{P}\cdot\dot{\vec{Q}}-\dot{\vec{P}}\cdot\vec{Q}}$ \\
\end{tabular}
\end{ruledtabular}
\end{table}
The Lie algebra obeyed by these vector fields is displayed in
TABLE~\ref{table:AdS}, where the quantities appearing in the table are
defined as follows
\begin{align}
\Omega^{u}&=\xi^{u}_{1}\dot{\xi}^{u}_2-\xi^{u}_{2}\dot{\xi}^{u}_1,
\label{eq:lambdaads}\\
\vec{\gamma}_{a}\bigl[\vec{P}\bigr]&=\xi^{u}_{a}\dot{\vec{P}}
-\frac{1}{2}\dot{\xi}^{u}_{a}\vec{P},\label{eq:gi} & a&=1,2,\\
\vec{P}_{ij}&=(P_{i}\delta^{k}_{j}-P_{j}\delta^{k}_{i})\bm{\partial_{k}},
\end{align}
and the structural constants ${c_{ij,kl}}^{mn}$ are those of standard
rotations,
\begin{align}
{c_{ij,kl}}^{mn}\bm{J}_{mn}=2\delta_{i[k}\bm{J}_{l]j}
-2\delta_{j[k}\bm{J}_{l]i}.
\end{align}

In the present case we have three different infinite-dimensional families
which make very complex and nontrivial the algebra of residual symmetries of
the AdS waves. The first family of generators \eqref{eq:xaads},
$\bm{\tilde{X}}_{\alpha}$, span an infinite-dimensional Abelian subalgebra.
We shall see later that it generates scalar supertranslations
\eqref{eq:t2ads}. An expansion in Fourier series of the function labeling
these generators,
\begin{equation}
\alpha(u)=\sum_{n=-\infty}^{\infty}a_{n}\left(-i\mathrm{e}^{-inu}\right),
\end{equation}
show that they are a linear combination,
\begin{equation}
\bm{\tilde{X}}_{\alpha}=\sum_{n=-\infty}^{\infty}a_{n}\bm{K}_{n},
\end{equation}
of the infinite commuting generators
\begin{equation}\label{eq:K_n}
\bm{K}_{n}=-i\mathrm{e}^{-inu}\left(\bm{\partial_{v}}
+2in\bm{\partial_{F}}\right).
\end{equation}
The Lie brackets of the family of generators \eqref{eq:xuads},
$\bm{\check{X}}_{\xi^{u}}$, among themselves have the same structure as the
infinite-dimensional subalgebras we have studied before both in the Collinson
\emph{Ansatz} and in the spherically symmetric one, and for which we have obtained
the Witt algebra. Following the same lines as in those cases, expanding the
function parametrizing the generator in Fourier series
\begin{equation}
\xi^{u}(u)=\sum_{n=-\infty}^{\infty}b_{n}\left(-i\mathrm{e}^{-inu}\right),
\end{equation}
the generator $\bm{\check{X}}_{\xi^{u}}$ becomes a linear superposition of
the infinite generators
\begin{align}
\bm{L}_{n}={}&-i\mathrm{e}^{-inu}\left[ \bm{\partial_{u}}
-\frac{in}{2}\left( y\bm{\partial_{y}}+x^{j}\bm{\partial_{j}}
-2F\bm{\partial_{F}}\right)\right.\nonumber\\
&\left.-\frac{n^2}{4}(y^2+\vec{x}^2)(\bm{\partial_{v}}
+2in\bm{\partial_{F}}) \right].\label{eq:lnads}
\end{align}
which also obey the Witt algebra \eqref{eq:alcon1}. It has been shown by
Ba\~nados, Chamblin and Gibbons \cite{Banados:1999tw} that the related
algebra of Noether charges associated to these symmetries acquires a central
extension in any dimension, generalizing the famous three-dimensional result
of Brown and Henneaux \cite{Brown:1986nw}. The last family of generators
\eqref{eq:kil2ads}, $\bm{X_}{\vec{P}}$, does not form an algebra by itself,
but it does in union with the Abelian family $\bm{\tilde{X}}_{\alpha}$. This
family generates vector supertranslations \eqref{eq:tx}. An expansion in
Fourier series of the vector
\begin{equation}
\vec{P}=\sum_{n=-\infty}^{\infty}\vec{c}_{n}(-i\mathrm{e}^{-inu}),
\end{equation}
allows us to write the vector field $\bm{X}_{\vec{P}}$ as a linear
combination,
\begin{equation}
\bm{X}_{\vec{P}}=\sum_{n=-\infty}^{\infty}(\vec{c}_{n})^{j}(\bm{M}_{n})_{j},
\end{equation}
of the infinite generators
\begin{equation}
(\bm{M}_{n})_{j}=-i\mathrm{e}^{-inu}\bm{\partial_{j}}-inx_{j}\bm{K}_n.
\label{eq:Mnj}
\end{equation}%
\begin{table*}[htbp!]
\caption{\label{table:AdSinf} The commutator table for AdS waves II:
infinite generators.}
\begin{ruledtabular}
\begin{tabular}{c|ccccc}
&$\bm{X}_1$ & $\bm{J}_{kl}$ & $\bm{K}_{m}$ &
$\bm{L}_{m}$ & $(\bm{M}_{m})_{l}$  \\
\noalign{\smallskip}\hline\noalign{\smallskip}
$\bm{X}_1$ & 0 & 0 & $-2\bm{K}_{m}$ & 0 & $-(\bm{M}_{m})_{l}$ \\
$\bm{J}_{ij}$ & 0 & ${c_{ij,kl}}^{mn}\bm{J}_{mn}$ & 0 & 0 &
$-2\delta_{l[i}(\bm{M}_{|m|})_{j]}$  \\
$\bm{K}_{n}$ & $2\bm{K}_{n}$ & 0 & 0 &
$n\bm{K}_{m+n}$ & 0 \\
$\bm{L}_{n}$   & 0 & 0 &
$-m\bm{K}_{m+n}$ &
$(n-m)\bm{L}_{n+m}$ &
$(n/2-m)(\bm{M}_{m+n})_{l}$ \\
$(\bm{M}_{n})_{i}$ & $(\bm{M}_{n})_{i}$ &
$2\delta_{i[k}(\bm{M}_{|n|})_{l]}$ & 0 &
$-(m/2-n)(\bm{M}_{n+m})_{i}$ &
$\delta_{il}(n-m)\bm{K}_{n+m}$ \\
\end{tabular}
\end{ruledtabular}
\end{table*}%
The Lie algebra can be written now in terms of the infinite generators
$\bm{L}_{n}$, $\bm{K}_{n}$, and $(\bm{M}_{n})_{j}$, which is done in
TABLE~\ref{table:AdSinf}. It is easier to identify in
TABLE~\ref{table:AdSinf} that the subalgebra spanned by the conformal
generators $\bm{L}_{n}$ and the scalar supertranslations $\bm{K}_{n}$ is the
semidirect sum of the Witt algebra and the loop algebra of $u(1)$. This
algebra and their central extension appear in the context of the Kerr/CFT and
WAdS/CFT correspondences and is used to compute the entropy of the involved
black holes
\cite{Guica:2008mu,Compere:2007in,Detournay:2012pc,Donnay:2015iia}. More
recently, it has been shown they appear also in the near horizon geometry of
stationary black holes \cite{Donnay:2015abr}.

We would like to emphasize that regardless of the complexity of the whole
infinite-dimensional subalgebra, it is possible to give a precise
interpretation for their infinite connected group as a well-defined subgroup
of the diffeomorphism group mapping AdS space to the AdS wave \emph{Ansatz}
\eqref{eq:AdSw}. In order to arrive at this interpretation we need to study
the finite version of the AdS waves residual symmetries following the methods
of App.~\ref{app:inv}. We start with the vector field \eqref{eq:x1ads}, which
gives rise to the anisotropic scaling
\begin{align}
\tilde{u}&=u, \qquad \tilde{v}=\lambda^{2}v, &
\tilde{y}&=\lambda{y}, & \vec{\tilde{x}}&=\lambda\vec{x},\nonumber\\
\tilde{F}&=\lambda^{2}F.\label{eq:t1ads}
\end{align}
The following vector field \eqref{eq:kil1ads} generates rotations
\begin{align}
\tilde{u}&=u, & \tilde{v}&=v, & \tilde{y}&=y, &
\tilde{x}^{i}&={\Lambda^{i}}_{j}x^{j},\nonumber\\
\tilde{F}&=F, \label{eq:rx}
\end{align}
where ${\Lambda^i}_j\in SO(D-3)$, i.e.\ it is a $(D-3)\times(D-3)$ orthogonal
matrix whose determinant is one. We emphasize these rotations are not
precisely isometries since the structural function $F$ depends in general on
the spatial coordinates and changes according to the diffeomorphisms rule
keeping invariant the value of the function at each point; this obviously
preserves the form of the \emph{Ansatz}. In the case of the generator
\eqref{eq:xaads}, the corresponding finite residual symmetry is
\begin{align}
\tilde{u}&=u,\qquad\tilde{v}=v+\alpha(u), & \tilde{y}&=y, &
\vec{\tilde{x}}&=\vec{x},\nonumber\\
\tilde{F}&=F-2\dot\alpha,\label{eq:t2ads}
\end{align}
we see that any scalar supertranslation along the null rays can be
compensated by another one in the function $F$ involving the derivative of
the original supertranslation. The residual symmetry related to generator
\eqref{eq:xuads} is a reparametrization of the retarded time which is
compensated accordingly
\begin{align}
\tilde{u}&=\tilde{u}(u), \qquad
\tilde{v}=v+\frac{1}{4}\frac{\mathrm{d}}{\mathrm{d}u}
\left(\ln\frac{\mathrm{d}\tilde{u}}{\mathrm{d}u}\right)
\left(y^2+\vec{x}^2\right),\nonumber\\
\tilde{y}&=\sqrt{\frac{\mathrm{d}\tilde{u}}{\mathrm{d}u}}y, \quad
\vec{\tilde{x}}=\sqrt{\frac{\mathrm{d}\tilde{u}}{\mathrm{d}u}}\vec{x},
\nonumber\\
\tilde{F}&=\left(\frac{\mathrm{d}\tilde{u}}{\mathrm{d}u}\right)^{-1}\!
\left[F+\left(\frac{\mathrm{d}\tilde{u}}{\mathrm{d}u}\right)^{1/2}
\frac{\mathrm{d}^{2}}
{\mathrm{d}u^{2}}\left(\frac{\mathrm{d}\tilde{u}}{\mathrm{d}u}\right)^{-1/2}
(y^2+\vec{x}^2)\right]\!.\label{eq:t3ads}
\end{align}
The details of their derivation can be followed step by step in
App.~\ref{app:AdS}. The integration of the generator \eqref{eq:kil2ads}
establishes the following finite transformation
\begin{align}
\tilde{u}&=u, \quad\!
\tilde{v}=v+\frac{\dot{\vec{P}}}2\!\cdot\!\left(2\vec{x}+\vec{P}\right)\!,
& \tilde{y}&=y, & \vec{\tilde{x}}&=\vec{x}+\vec{P}(u),\nonumber\\
\tilde{F}&=F-\ddot{\vec{P}}\!\cdot\!\left(2\vec{x}+\vec{P}\right)\!,
\label{eq:tx}
\end{align}
where this time a vector supertranslation along the spatial coordinates
$\vec{x}$ can be suitably compensated for.

The most general connected residual symmetry of AdS waves is the composition
of the above transformations and can be written as
\begin{subequations}\label{eq:frsAdSw}
\begin{align}
\tilde{u}={}&\int{\frac{\mathrm{d}u}{f^2}},\nonumber\\
\tilde{v}={}&\lambda^2\Biggl\{v
-\frac{1}{2}\frac{\dot{f}}{f}\left(y^2+\vec{x}^{2}\right)
+f\frac{\mathrm{d}}{\mathrm{d}u}
\left(f^{-1}\vec{P}\right)\cdot\vec{x}
\nonumber\\
&+\frac{1}{2}\int{\mathrm{d}u\left[F_0+\dot{\vec{P}}^{2}
-\dot{f}\frac{\mathrm{d}}{\mathrm{d}u}
\left(f^{-1}\vec{P}^2\right)\right]}\Biggr\},\nonumber\\
\tilde{y}={}&\frac{\lambda}{f}\,y, \qquad
\vec{\tilde{x}}=\frac{\lambda}{f}\,
\tensor{\Lambda}\cdot\left(\vec{x}+\vec{P}\right),
\nonumber\\
\tilde{F}={}&(\lambda f)^2\left[F-F_2\left(y^2+\vec{x}^{2}\right)
-\vec{F}_1\cdot\vec{x}-F_0\right],
\label{eq:tgads}
\end{align}
where $f$, $F_0$ and $\vec{P}$ are arbitrary functions of the retarded time
$u$ determining the functions
\begin{equation}\label{eq:F_21}
F_2=-\frac{\ddot{f}}{f}, \qquad \vec{F}_1=2(\ddot{\vec{P}}+\vec{P}F_2),
\end{equation}
\end{subequations}
and the involved parameters are the scaling constant $\lambda$ and the matrix
$\tensor{\Lambda}\in SO(D-3)$. Except for the anisotropic scaling, this
transformation becomes the one already known in the literature for $D=4$
\cite{Siklos:1985} and $D=3$ \cite{AyonBeato:2005qq}, since no rotation
exists for these dimensions. The resulting transformation is the
infinite-dimensional subgroup of the connected group of residual symmetries,
which can be understood in lower and higher dimension according to the
following lines. It is well-known that a vanishing structural constant,
$F=0$, is just the AdS spacetime in Poincare coordinates. However, as was
first noticed by Siklos \cite{Siklos:1985}, this is only the minimal way to
represent AdS spacetime with the AdS wave \emph{Ansatz}. Imposing that the AdS waves
have constant curvature (Weyl and traceless Ricci tensors vanish), i.e.\ that
they are locally equivalent to AdS, the following expression for the
structural function is obtained
\begin{equation}\label{eq:AdS}
F_{\textrm{AdS}}=F_2\left(y^2+\vec{x}^{2}\right)
+\vec{F}_1\cdot\vec{x}+F_0,
\end{equation}
where here $F_0$, $\vec{F}_1$ and $F_2$ are arbitrary functions of the
retarded time $u$. As is proved at the end of App.~\ref{app:AdS}, this is the
most general way to express AdS spacetime with the AdS wave \emph{Ansatz}
\eqref{eq:AdSw} in any dimension and it must be locally equivalent to the
metric with $F=0$. It is straightforward to check that the local
transformation reducing the above expression for the structural function to
the vanishing one is precisely the infinite-dimensional subgroup of the
connected group of residual symmetries contained in \eqref{eq:frsAdSw} for
$\lambda=1$ and $\tensor{\Lambda}=\openone$. The only requirement is to
choose the functions in the transformation to coincide with the ones in
\eqref{eq:AdS}, which imposes differential equations for the functions $f$
and $\vec{P}$ via the relations \eqref{eq:F_21}. Consequently, for a given
AdS wave any quadratic, linear and homogeneous dependencies on the front-wave
coordinates as the ones appearing in \eqref{eq:AdS}, are reminiscences of the
freedom to express AdS spacetime within this \emph{Ansatz} and can be consistently
eliminated. This is the interpretation of the infinite-dimensional sector of
residual symmetries of the AdS wave \emph{Ansatz}.

\section{\label{sec:Papa}The Papapetrou \emph{Ansatz}}

In this section we study the popular \emph{Ansatz} of Papapetrou
\cite{Papapetrou:1953zz} which also describes circular stationary
axisymmetric spacetimes, as the initial Collinson \emph{Ansatz} \eqref{eq:coll}, and
whose metric is given by
\begin{equation}\label{eq:ppp}
\bm{\mathrm{d}s^{2}}=-\frac{\rho^{2}}{X}\bm{\mathrm{d}t^{2}}
+X\left(\bm{\mathrm{d}\varphi}+A\bm{\mathrm{d}t}\right)^{2}
+\frac{e^{2h}}{X}\left(\bm{\mathrm{d}\rho^{2}}+\bm{\mathrm{d}z^{2}}\right),
\end{equation}
where $X$, $h$ and $A$ are functions of the spatial coordinates $\rho$ and
$z$ only. The main difference of the Papapetrou \emph{Ansatz} with respect to the
one of Collinson is that the conformal freedom of the latter is fixed in the
former choosing the so-called Weyl coordinates \cite{HeM}. These coordinates
exploit the fact that in vacuum and electro-vacuum, one of the gravitational
potentials becomes harmonic and can be incorporated in the conformal
transformation to play the role of the real or imaginary part of the spatial
complex coordinates of the Collinson \emph{Ansatz}. Concretely, this potential is
identified with the spatial coordinate $\rho$ in \eqref{eq:ppp}. As result,
the Papapetrou \emph{Ansatz} has one structural function less than in the Collinson case
and this is the starting point of the standard approach to study integrable
circular stationary axisymmetric systems in general relativity and which
results in the so-called Ernst equations \cite{Ernst:1967wx}. The jet
space coordinates for the Papapetrou \emph{Ansatz} are
$x^{\mu}=\left(t,\varphi,\rho,z\right)$ and $u^{I}=\left(X,A,h\right)$. The
generators of residual symmetries that we found in App.~\ref{app:Papa} using
the criterion \eqref{eq:crit} are shown below
\begin{subequations}\label{eq:gensppp}
\begin{align}
\bm{X}_{1}&=t\bm{\partial_{t}}+\varphi\bm{\partial_\varphi}
-2\rho\bm{\partial_{\rho}}-2z\bm{\partial_{z}}
-2X\bm{\partial_{X}}+\bm{\partial_{h}},\label{eq:x1ppp}\\
\bm{X}_{2}&=t\bm{\partial_{t}}-\varphi\bm{\partial_\varphi}
+2X\bm{\partial_{X}}-2A\bm{\partial_{A}}+\bm{\partial_{h}},\label{eq:x2ppp}\\
\bm{X}_{3}&=\varphi\bm{\partial_{t}}-2AX\bm{\partial_{X}}
+\left(A^{2}+{\rho^{2}}/{X^{2}}\right)\bm{\partial_{A}}
-A\bm{\partial_{h}},\label{eq:x3ppp}\\
\bm{X}_{4}&=t\bm{\partial_{\varphi}}-\bm{\partial_{A}},\label{eq:x4ppp}\\
\bm{X}_{5}&=\bm{\partial_{z}},\label{eq:x5ppp}\\
\bm{k}&=\bm{\partial_{t}},\label{eq:kil1ppp}\\
\bm{m}&=\bm{\partial_{\varphi}}\label{eq:kil2ppp}.
\end{align}
\end{subequations}
We recognize in the generators (\ref{eq:kil1ppp}) and (\ref{eq:kil2ppp}) to
the stationary and axisymmetric Killing vectors, the rest are generators of
genuine residual symmetries and they together form the Lie algebra exhibited
in TABLE~\ref{table:papapetrou}.%
\begin{table}[hbtp!]
\caption{\label{table:papapetrou} The commutator table for the Papapetrou \emph{Ansatz}.}
\begin{ruledtabular}
\begin{tabular}{c|ccccccc}
& $\bm{X}_1$ & $\bm{X}_2$ & $\bm{X}_3$ & $\bm{X}_4$ & $\bm{X}_5$ & $\bm{k}$ &
$\bm{m}$ \\
\noalign{\smallskip}\hline\noalign{\smallskip}
$\bm{X}_1$ & 0 & 0 & 0 & 0 & $2\bm{X}_5$ & $-\bm{k}$ & $-\bm{m}$ \\
$\bm{X}_2$ & 0 & 0 & $-2\bm{X}_3$ & $2\bm{X}_4$ & 0 & $-\bm{k}$ &
$\bm{m}$ \\
$\bm{X}_3$ & 0 & $2\bm{X}_3$ & 0 & $-\bm{X_{2}}$ & 0 & 0 & $-\bm{k}$ \\
$\bm{X}_4$ & 0 & $-2\bm{X}_4$ & $\bm{X}_2$ & 0 & 0 &
$-\bm{m}$ & 0 \\
$\bm{X}_5$ & $-2\bm{X}_5$ & 0 & 0 & 0 & 0 & 0 & 0 \\
$\bm{k}$ & $\bm{k}$ & $\bm{k}$ & 0 & $\bm{m}$ & 0 & 0 & 0 \\
$\bm{m}$ & $\bm{m}$ & $-\bm{m}$ & $\bm{k}$ & 0 & 0 & 0 & 0 \\
\end{tabular}
\end{ruledtabular}
\end{table}

The finite transformations generated by these vector fields are easily
obtained using the methods of App.~\ref{app:inv}. After integrating the
generator \eqref{eq:x1ppp} we obtain
\begin{align}
\tilde{t}&=\lambda t, & \tilde{\varphi}&=\lambda\varphi, &
\tilde{\rho}&=\lambda^{-2}\rho, & \tilde{z}&=\lambda^{-2}z,\nonumber\\
\widetilde{X}&=\lambda^{-2} X, & \widetilde{A}&=A, &
\mathrm{e}^{2\tilde{h}}&=\lambda^{2}\mathrm{e}^{2h}.\label{eq:t1ppp}
\end{align}
We see that scaling Killing coordinates is compensated for by other scalings in
both the remaining spatial coordinates and the structural functions. This
transformation and their generator \eqref{eq:x1ppp} are similar to the
Killing scaling of the Collinson \emph{Ansatz}, see Eqs.~\eqref{eq:x1col} and
\eqref{eq:t1col}. The finite transformation we found in the case of generator
(\ref{eq:x2ppp}) is the following
\begin{align}
\tilde{t}&=\lambda{t}, & \tilde{\varphi}&=\lambda^{-1}\varphi, &
\tilde{\rho}&=\rho, & \tilde{z}&=z,\nonumber\\
\widetilde{X}&=\lambda^{2}X, & \widetilde{A}&=\lambda^{-2}A, &
\mathrm{e}^{2\tilde{h}}&=\lambda^{2}\mathrm{e}^{2h}.\label{eq:t2ppp}
\end{align}
Therefore, scaling Killing coordinates inversely can be also compensated for with
appropriated scalings, this time from the structural functions only. This is
the analog of the inverse Killing scaling of the Collinson \emph{Ansatz} described
in Eqs.~\eqref{eq:x2col} and \eqref{eq:t2col}. In the case of the generator
\eqref{eq:x3ppp} we have the following transformation
\begin{align}
\tilde{t}&=t+\varepsilon\varphi,\quad\tilde{\varphi}=\varphi,\quad
\tilde{\rho}=\rho,\quad\tilde{z}=z,\frac{}{}\nonumber\\
\widetilde{X}&=X\left[1-2{\varepsilon}A+\varepsilon^{2}\left(A^{2}
-{\rho^{2}}/{X^{2}}\right)\right],\frac{}{}\nonumber\\
\widetilde{A}&=\frac{A-\varepsilon\left(A^{2}
-{\rho^{2}}/{X^{2}}\right)}{1-2{\varepsilon}A
+\varepsilon^{2}\left(A^{2}-{\rho^{2}}/{X^{2}}\right)},\nonumber \\
\mathrm{e}^{2\widetilde{h}}&=\mathrm{e}^{2h}\left[1-2{\varepsilon}A
+\varepsilon^{2}\left(A^{2}-{\rho^{2}}/{X^{2}}\right)\right].\frac{}{}
\label{eq:t4ppp}
\end{align}
In other words, rotating time in the Killing plane is compensated for by
redefining the structural functions with a sort of special conformal
transformation. In turn, integration of the vector field \eqref{eq:x4ppp}
gives a rotation of the angle in the Killing plane,
\begin{align}
\tilde{t}&=t, & \tilde{\varphi}&=\varphi+\varepsilon t, &
\tilde{\rho}&=\rho, & \tilde{z}&=z,\nonumber\\
\widetilde{X}&=X, & \widetilde{A}&=A-\varepsilon, &
\mathrm{e}^{2\tilde{h}}&=\mathrm{e}^{2h},\label{eq:t3ppp}
\end{align}
which is compensated for with a translation. These last two Killing rotations are
equivalent to those appearing for the Collinson \emph{Ansatz}, Eqs.~\eqref{eq:t3col}
and \eqref{eq:t4col}, with generators \eqref{eq:x3col} and \eqref{eq:x4col}.
The simplest residual symmetry for the Papapetrou \emph{Ansatz} comes from the
vector field \eqref{eq:x5ppp} and is a translation on the spatial coordinate
$z$. This is not an isometry since the structural functions depend in general
on this coordinate, but they will change as functions do under
diffeomorphisms which preserve the form of the \emph{Ansatz}. This $z$ translation
is the residual symmetry left after breaking the conformal freedom of the
Collinson \emph{Ansatz} by fixing Weyl coordinates choosing one of the gravitational
potentials as the coordinate $\rho$.

The composition of the latter transformations leads to the most general
connected residual symmetry admitted by the Papapetrou metric (\ref{eq:ppp}),
\begin{align}
\tilde{t}&=\alpha{t}+\beta\varphi, \qquad\, 
\tilde{\varphi}=\gamma{t}+\delta\varphi, \frac{}{}\nonumber\\
\tilde{\rho}&=\frac{\rho}{\alpha\delta-\beta\gamma}, \qquad
\tilde{z}=\frac{z}{\alpha\delta-\beta\gamma}+\epsilon,\nonumber\\
\widetilde{X}&=X\frac{\alpha^2-2\alpha\beta{A}+\beta^{2}(A^2-\rho^2/X^2)}
{(\alpha\delta-\beta\gamma)^2},\nonumber\\
\widetilde{A}&=
\frac{(\alpha\delta+\beta\gamma)A-\beta\delta(A^2-\rho^2/X^2)-\alpha\gamma}
{\alpha^2-2\alpha\beta{A}+\beta^{2}(A^2-\rho^2/X^2)},\nonumber\\
\mathrm{e}^{2\tilde{h}}&=\mathrm{e}^{2h}
\left[\alpha^2-2\alpha\beta{A}+\beta^{2}(A^2-\rho^2/X^2)\right],
\label{eq:tgenppp}
\end{align}
where $\alpha\delta-\beta\gamma$ is a positive quantity. A very well-known residual symmetry of the Papapetrou \emph{Ansatz} is the one
linking the so-called conjugate potentials introduced by Chandrasekhar.
Between other applications it allows us to derive the famous Kerr metric
\cite{Kerr:1963ud} from an unphysical trivial solution to the Ernst equations
\cite{HeM}. It is easy to check that taking the following values for the
parameters of transformation (\ref{eq:tgenppp}),
\begin{equation}\label{eq:paramppp}
\alpha=\delta=\tau=0, \quad \beta=-\gamma=1,
\end{equation}
one recovers the relation defining conjugate potentials \cite{HeM},
\begin{align}
\tilde{t}&=\varphi,\quad\tilde{\varphi}=-t,\quad\tilde{\rho}=\rho,\quad
\tilde{z}=z,\nonumber\\
\widetilde{X}&=X\left(A^{2}-\frac{\rho^{2}}{X^{2}}\right),\nonumber\\
\widetilde{A}&=-A\left(A^{2}-\frac{\rho^{2}}{X^{2}}\right)^{-1},\nonumber\\
e^{2\tilde{h}}&=e^{2h}\left(A^{2}-\frac{\rho^{2}}{X^{2}}\right).
\label{eq:contran}
\end{align}

\section{\label{sec:noncircColl}The noncircular Collinson \emph{Ansatz}}

The so-called Collinson theorem we mention at the beginning of
Sec.~\ref{sec:Coll} was first established for circular stationary
axisymmetric spacetimes for which the metric is block diagonal with one block
related with the Killing directions and the other with the remaining spatial
sector \cite{Collinson:1976,Garcia:2002gj}. It was later extended to general
stationary axisymmetric spacetimes by considering also noncircular
contributions \cite{AyonBeato:2005yx}. This was possible due to the following
generalization of the Collinson \emph{Ansatz}
\begin{align}
\bm{\mathrm{d}s^{2}}={}&e^{-2Q}\biggl(
-\frac{1}{a+b}\left(\bm{\mathrm{d}\tau}+a\bm{\mathrm{d}\sigma}
+M\bm{\mathrm{d}y}\right)\nonumber\\
&\times\left(\bm{\mathrm{d}\tau}-b\,\bm{\mathrm{d}\sigma}
-N\bm{\mathrm{d}y}\right)
+e^{-2P}(\bm{\mathrm{d}x^2}+\bm{\mathrm{d}y^2})\biggr),\quad\quad
\label{eq:collgen2}
\end{align}
describing the most general stationary axisymmetric spacetimes and where the
structural functions include now the noncircular contributions $M$ and $N$,
and all the functions depend exclusively of the spatial coordinates $(x,y)$
\cite{AyonBeato:2005yx}. In this case the jet space coordinates
$z^{A}=\left(x^{\mu}, u^{I}\right)$ are given by
$x^{\mu}=\left(\tau,\sigma,x,y\right)$ and $u^{I}=\left(a,b,P,Q,M,N\right)$.
The infinitesimal generators of residual symmetries found in
App.~\ref{app:noncircColl} for the generalization \eqref{eq:collgen2} are
listed below
\begin{subequations}\label{eq:genscolgen}
\begin{align}
\bm{X}_{1}&=\tau\bm{\partial_{\tau}}+\sigma\bm{\partial_{\sigma}}
-\bm{\partial_{P}}+\bm{\partial_{Q}}
+M\bm{\partial_{M}}+N\bm{\partial_{N}},\label{eq:x1colgen}\\
\bm{X_}{2}&=\tau\bm{\partial_{\tau}}-\sigma\bm{\partial_{\sigma}}
+2a\bm{\partial_{a}}+2b\bm{\partial_{b}}
+M\bm{\partial_{M}}+N\bm{\partial_{N}},\label{eq:x2colgen}\\
\bm{X_}{3}&=\sigma\bm{\partial_{\tau}}-\bm{\partial_{a}}+\bm{\partial_{b}},
\label{eq:x3colgen}\\
\bm{X_}{4}&=\tau\bm{\partial_{\sigma}}+a^{2}\bm{\partial_{a}}
-b^{2}\bm{\partial_{b}}+Ma\bm{\partial_{M}}-Nb\bm{\partial_{N}},
\label{eq:x4colgen}\\
\bm{X_}{5}&=x\bm{\partial_{x}}+y\bm{\partial_{y}}+\bm{\partial_{P}}
-M\bm{\partial_{M}}-N\bm{\partial_{N}},\label{eq:x5colgen}\\
\bm{X_}{6}&=\bm{\partial_{x}},\label{eq:x6colgen}\\
\bm{X_}{7}&=\bm{\partial_{y}},\label{eq:x7colgen}\\
\bm{X}_{F_{1}}&=F_{1}(y)\bm{\partial_{\sigma}}
-F_{1}^{\prime}(a\bm{\partial_{M}}+b\bm{\partial_{N}}),\label{eq:xf1colgen}\\
\bm{\hat{X}}_{F_{2}}&=F_{2}(y)\bm{\partial_{\tau}}
-F_{2}^{\prime}(\bm{\partial_{M}}-\bm{\partial_{N}}).\label{eq:xf2colgen}
\end{align}
\end{subequations}
Note that the Killing vectors of the metric are a particular case of the
generators \eqref{eq:xf1colgen} and \eqref{eq:xf2colgen} when the involved
functions are constant, in other cases they generalize to two commuting
infinite-dimensional Abelian Lie subalgebras.%
\begin{table*}[htbp!]
\caption{\label{table:noncColl} The commutator table for the noncircular
Collinson \emph{Ansatz}.}
\begin{ruledtabular}
\begin{tabular}{c|ccccccccc}
& $\bm{X}_1$ & $\bm{X}_2$ & $\bm{X}_3$ & $\bm{X}_4$ & $\bm{X}_5$ & $\bm{X}_6$
& $\bm{X}_7$ & $\bm{X}_{F_{1}}$ & $\bm{\hat{X}}_{F_{2}}$ \\
\noalign{\smallskip}\hline\noalign{\smallskip}
$\bm{X}_1$ & 0 & 0 & 0 & 0 & 0 & 0 & 0 & $-\bm{X}_{F_{1}}$ &
$-\bm{\hat{X}}_{F_{2}}$ \\
$\bm{X}_2$ & 0 & 0 & $-2\bm{X}_3$ & $2\bm{X}_4$ & 0 & 0 & 0 &
$\bm{X}_{F_{1}}$ & $-\bm{\hat{X}}_{F_{2}}$ \\
$\bm{X}_3$ & 0 & $2\bm{X}_3$ & 0 & $-\bm{X}_2$ & 0 & 0 & 0 &
$-\bm{\hat{X}}_{F_{1}}$ & 0 \\
$\bm{X}_4$ & 0 & $-2\bm{X}_4$ & $\bm{X}_2$ & 0 & 0 & 0 & 0 & 0 &
$-\bm{X}_{F_{2}}$ \\
$\bm{X}_5$ & 0 & 0 & 0 & 0 & 0 & $-\bm{X}_6$ & $-\bm{X}_7$ &
$\bm{X}_{yF^{\prime}_{1}}$ & $\bm{\check{X}}_{yF^{\prime}_{2}}$ \\
$\bm{X}_6$ & 0 & 0 & 0 & 0 & $\bm{X}_6$ & 0 & 0 & 0 & 0 \\
$\bm{X}_7$ & 0 & 0 & 0 & 0 & $\bm{X}_7$ & 0 & 0 & 0 & 0 \\
$\bm{X}_{G_{1}}$ & $\bm{X}_{G_{1}}$ & $-\bm{X}_{G_{1}}$ &
$\bm{\hat{X}}_{G_{1}}$ & 0 & $-\bm{X}_{yG^{\prime}_{1}}$ & 0 & 0 & 0 & 0 \\
$\bm{\hat{X}}_{G_{2}}$ & $\bm{\hat{X}}_{G_{2}}$ & $\bm{\hat{X}}_{G_{2}}$ & 0 &
$\bm{X}_{G_{2}}$ & $-\bm{\check{X}}_{yG^{\prime}_{2}}$ & 0 & 0 & 0 & 0 \\
\end{tabular}
\end{ruledtabular}
\end{table*}
The rest of the generators \eqref{eq:x1colgen}-\eqref{eq:x7colgen} form a
seven-dimensional Lie subalgebra, and all them together give the algebra
observed in TABLE~\ref{table:noncColl}.

The generators \eqref{eq:x1colgen}-\eqref{eq:x4colgen} are just those
appearing for the circular Collinson \emph{Ansatz}
\eqref{eq:x1col}-\eqref{eq:x4col}, but modified by noncircular contributions.
An exception is the generator \eqref{eq:x3colgen} which is exactly the same
for both spacetimes, as is appreciated in Eq.~\eqref{eq:x3col}. The vector
fields \eqref{eq:x5colgen}, \eqref{eq:x6colgen} and \eqref{eq:x7colgen}
correspond to the cases where the conformal generator \eqref{eq:xxicol}
becomes $\xi^{z}(z)=z=x+{i}y$, $\xi^{z}(z)=1$ and $\xi^{z}(z)=i$,
respectively, where the first case is additionally supplemented by
noncircular contributions. In other words, the infinite conformal freedom
enjoyed by the circular Collinson \emph{Ansatz} breaks in these three generators,
which is a consequence of the presence of noncircular terms. Conversely, as
we check later, the standard translations symmetries of the circular Killing
vectors \eqref{eq:kill1col} and \eqref{eq:kill2col} enhance to
supertranslations \eqref{eq:xf1colgen} and \eqref{eq:xf2colgen} depending on
the direction along which the circularity is lost.

We study now how the finite transformations are extended to include
noncircular components using again the methods of App.~\ref{app:inv}. The
finite residual symmetry obtained after integrating generator
(\ref{eq:x1colgen}) is
\begin{align}
\tilde{\tau}&=\lambda{\tau}, & \tilde{\sigma}&=\lambda\sigma, &
\tilde{x}&=x, & \tilde{y}&=y,\nonumber\\
\tilde{a}&=a, & \tilde{b}&=b, & \tilde{P}&=P-\ln\lambda, &
\tilde{Q}&=Q+\ln\lambda, \nonumber\\
\tilde{M}&=\lambda{M}, &
\tilde{N}&=\lambda{N},\label{eq:t1colgen}
\end{align}
which is just the Killing scaling \eqref{eq:t1col} trivially extended to the
noncircular portions. Something similar happens with generator
\eqref{eq:x2colgen} with regard to the anisotropic Killing scaling
\eqref{eq:t2col}
\begin{align}
\tilde{\tau}&=\lambda\tau, & \tilde{\sigma}&=\lambda^{-1}\sigma, &
\tilde{x}&=x, & \tilde{y}&=y,\nonumber\\
\tilde{a}&=\lambda^{2}a, & \tilde{b}&=\lambda^{2}b, &
\tilde{P}&=P, & \tilde{Q}&=Q,\nonumber\\
\tilde{M}&=\lambda{M}, & \tilde{N}&=\lambda{N}.
\label{eq:t2colgen}
\end{align}
Since the generator \eqref{eq:x3colgen} is the same as in the circular case
\eqref{eq:x3col}, their finite residual symmetry is the time rotation
\eqref{eq:t3col} used as example at the beginning of the work and with no
effect along the noncircular contributions. In opposition, the angular
rotation generated by the vector field \eqref{eq:x4colgen},
\begin{align}
\tilde{\tau}&=\tau, & \tilde{\sigma}&=\sigma+\varepsilon\tau, &
\tilde{x}&=x, & \tilde{y}&=y, \nonumber\\
\tilde{a}&=\frac{a}{1-\varepsilon a}, &
\tilde{b}&=\frac{b}{1+\varepsilon b}, &
\tilde{P}&=P, & \tilde{Q}&=Q,\nonumber\\
\tilde{M}&=\frac{M}{1-\varepsilon a}, &
\tilde{N}&=\frac{N}{1+\varepsilon b},\label{eq:t4colgen}
\end{align}
nontrivially extends the circular one \eqref{eq:t4col}, since it needs to be
also compensated for by special conformal transformations of the noncircular
structural functions. The residual transformation one gets after
exponentiation of the vector field \eqref{eq:x5colgen} is the following
spatial scaling
\begin{align}
\tilde{\tau}&=\tau, & \tilde{\sigma}&=\sigma, & \tilde{x}&=\lambda{x}, &
\tilde{y}&=\lambda{y},\nonumber\\
\tilde{a}&=a, & \tilde{b}&=b, &
\tilde{P}&=P+\ln\lambda, & \tilde{Q}&=Q\nonumber\\
\tilde{M}&=\lambda^{-1}M, & \tilde{N}&=\lambda^{-1}N,\label{eq:t3colgen}
\end{align}
which is compensated for by a translation in $P$ and inverse scalings in $N$ and
$M$. The conformal symmetry \eqref{eq:t5col} breaks to this spatial scaling
and to the spatial translations \eqref{eq:x6colgen} and \eqref{eq:x7colgen}.
We emphasize these last two are not precisely isometries, since in general all
the structural functions depend on the spatial coordinates and will change
according to the diffeomorphism rule for functions, which preserve the form
of the \emph{Ansatz}. The generator \eqref{eq:xf1colgen} leads to the following
finite residual symmetry
\begin{alignat}{2}
\tilde{\tau}&=\tau, \qquad \tilde{\sigma}=\sigma+F_{1}(y), & \qquad\quad
\tilde{x}&=x, \qquad\,\, \tilde{y}=y,\nonumber\\
\tilde{a}&=a, \qquad\, \tilde{b}=b, &
\tilde{P}&=P, \qquad \tilde{Q}=Q,\nonumber\\
\tilde{M}&=M-aF_{1}^{\prime}, & \tilde{N}&=N-bF_{1}^{\prime},
\label{eq:t4colgen2}
\end{alignat}
saying that the circular translation isommetry along the angle is improved to
a supertranslation whose local dependence is determined by the spatial
direction where the noncircularity appears and which is compensated for by
superrotating the noncircular structural functions $M$ and $N$ in the planes
$(a,M)$ and $(b,N)$, respectively. The vector field \eqref{eq:xf2colgen}
enhances the time translation isometry of the circular case to a similar
supertranslation
\begin{align}
\tilde{\tau}&=\tau+F_{2}(y), & \tilde{\sigma}&=\sigma, \qquad
\tilde{x}=x, & \tilde{y}&=y,\nonumber\\
\tilde{a}&=a, & \tilde{b}&=b, \qquad \tilde{P}=P, & \tilde{Q}&=Q\nonumber\\
\tilde{M}&=M-F_{2}^{\prime}, & \tilde{N}&=N+F_{2}^{\prime},
\label{eq:t5colgen}
\end{align}
but compensated for this time with supertranslations along the noncircular
structural functions.

Finally, we would like to comment that there is a more symmetrical version of
the noncircular \emph{Ansatz} where the diffeomorphism gauge freedom is not
enterally fixed, and that involves to consider $M$ and $N$ as arbitrary
complex functions \cite{AyonBeato:2005yx}. This has the advantage of
preserving the conformal invariance of the circular case. However, as is
explicitly proved here this huge and useful symmetry is lost after gauge
fixing.

\section{\label{sec:conclu}Conclusions}

In this work we have analyzed how to unambiguously define the residual
symmetries of a gravitational \emph{Ansatz}. The intuitive conditions that spacetime
diffeomorphisms must be compensated for with redefinitions of the structural
functions determining the \emph{Ansatz} have a very nice and geometrical
interpretation in terms of general diffeomorphisms on jet space, where
both spacetimes coordinates and the functions are independent variables.
Consequently, these conditions are interpreted such that the generators of
residual symmetries are jet space Killing fields of the trivial
lifting of the metric. However, it has the disadvantage of including
excessively general transformations that only preserve the \emph{Ansatz} formally by
changing the original dependencies of the structural functions. Studying the
transformations of the structural functions derivatives, we provide the
complementary conditions needed to avoid this situation and obtain residual
symmetries in a strict sense. The supplemental conditions make the
integration of the system of partial differential equations determining the
generators easier. In fact, the full criterion provides an effective computational
procedure for finding all the residual symmetries of almost any gravitational
\emph{Ansatz} of interest. We describe also how to integrate the resulting
infinitesimal generators to obtain the related finite transformations by
means of the efficient method of differential invariants.

We apply the criterion to study five different gravitational \emph{Ans\"atze}. We start
with the case of the Collinson \emph{Ansatz} describing circular stationary
axisymmetric spacetimes and we recover all the residual symmetries previously
found by Collinson himself. Apart from the isometries, the finite dimensional
subalgebra includes two Killing scalings and two Killing rotations
appropriately compensated for by scalings, translations and special conformal
transformations on the structural functions. There is also an
infinite-dimensional subalgebra which is just a conformal symmetry on the
spatial plane orthogonal to the Killing vectors. Using a complex Laurent
series we characterize the concrete infinite generators spanning the two
related copies of the Witt algebra.

Later we study the spherically symmetric \emph{Ansatz} in its static and
time-dependent versions and previous to completely fixing the gauge on the
radial variable. In the static case in addition to the isometries there is a
time scaling and consequently a radial reparameterization as residual
symmetries. In the time-dependent case we have the same radial
reparameterization and loosing the stationary isometry enhances the above
time scaling to a time reparameterization. We identify, by means of a Fourier
expansion this time, the infinite generators providing also in this example
two copies of the Witt algebra characterizing the conformal symmetry now
enjoyed by the Lorentzian fibers orthogonal to the spheres.

We also generalize the residual symmetries already known for lower
dimensional AdS waves to higher dimensions. The resulting
infinite-dimensional subalgebra has a very complex structure formed by three
infinite-dimensional families. One of them form a Witt algebra by itself and
has been the base to show that the appearance of a central extension
associated with their Noether charges is not exclusive of three dimensions.
The other two families are scalar and vector supertranslations forming a more
complex algebra. Independently of this complexity we manage to understand the
precise meaning of the infinite connected group. It is just the well-defined
subgroup of the diffeomorphism group encoding the freedom to represent the
AdS space by means of the AdS wave \emph{Ansatz}.

We study another famous example of an \emph{Ansatz} describing a circular stationary
axisymmetric spacetime, which is the Papapetrou \emph{Ansatz}. It is restricted by
the election of the so-called Weyl coordinates, which exist when one of the
gravitational potentials is harmonic and can be choosen as one of the spatial
coordinates. A situation occurring for vacuum and electro-vacuum spacetimes.
Obviously, this election breaks the conformal symmetry of the circular
stationary axisymmetric spacetimes explicitly manifest in the Collinson
\emph{Ansatz}. We found the same Killing scalings and rotations than the Collinson
case together with the isometries. We also observe that the previous
conformal freedom breaks to a single translation of the other spatial
coordinate not identified with one of the gravitational potentials. In other
words, Weyl coordinates are undetermined modulo this translation.
Additionally, from the general connected group of resulting residual
symmetries we identify the precise point defining the relation between
conjugate potentials introduced by Chandrasekhar. They are well-known by
allowing us to derive the famous Kerr metric from an unphysical trivial solution
of the Ernst equations.

Finally, we also explore the inclusion of noncircular contributions in the
Collinson \emph{Ansatz}. There are two immediate effects of losing circularity, the
first is that again the conformal symmetry is broken, in this case to spatial
translations and a scaling. The second is that the standard translations
symmetries of the circular Killing vectors enhance to supertranslations which
depend on the direction along which the circularity is lost.

An immediate generalization of our work is to consider gravitational
\emph{Ans{\"a}tze} that also depend on derivatives of the structural functions.
There are interesting examples of this kind in the literature, we will report
on these examples in the future. Another extension which can be very useful
is to try to include in the unifying perspective of the present approach the
study of the asymptotic symmetries of spacetimes, which are nothing than a
very particular kind of residual symmetries.

Last but not least, an issue that must be inevitably addressed is the further
development of these methods to include matter fields. In the case of a gauge
field one should look for diffemorphisms and gauge transformations that
together are compensated for by redefinitions of the structural functions
defining the \emph{Ansatz} for the gauge field. However, it is enough to study the
residual symmetries of gauge invariant quantities as the field strength in
the Abelian case or the energy-momentum tensor in the non-Abelian one, the latter simply being
the quantity defining the coupling to the gravitational field.
The additional criterion must reduce to the vanishing of the \emph{jet space}
Lie derivative of these gauge invariant quantities along generators prolonged
in the new structural directions.

\begin{acknowledgments}
We are thankful to F.~Canfora, A.~Garc\'{\i}a, G.~Giribet, M.~Hassa\"{\i}ne,
T.~Matos and C.~Troessaert for enlightening and helpful discussions. This
work has been funded by grants No. 175993 and No. 178346 from CONACyT, together with
grants No. 1121031, No. 1130423 and No. 1141073 from FONDECYT. E. A. B. was partially
supported by the Programa Atracci\'{o}n de Capital Humano Avanzado del
Extranjero, MEC from CONICYT. G. V. R. was partially supported by the Programa
de Becas Mixtas from CONACyT and by the Plataforma de Movilidad
Estudiantil Alianza del Pac\'{\i}fico of AGCI.
\end{acknowledgments}

\appendix

\section{Differential invariants method\label{app:inv}}

Any Lie point transformation \eqref{eq:Lie_pt} defines an infinitesimal
generator \eqref{eq:genjs} on jet space
$z^{A}=\left(x^{\mu},u^{I},\ldots\right)$. Conversely, given the
infinitesimal generator one can find the associated finite transformations,
$\tilde{z}^{B}=\tilde{z}^{B}(z^{A};\varepsilon)$, which are defined as the
solution to the dynamical system
\begin{equation}\label{eq:dyn}
\frac{\mathrm{d}\tilde{z}^{B}(\varepsilon)}{\mathrm{d}\varepsilon}
=X^{B}(\tilde{z}^{A}(\varepsilon)), \qquad \tilde{z}^{B}(0)=z^{B},
\end{equation}
where the starting coordinates play the role of the initial conditions. In
this appendix we review how to integrate this system and find the finite
transformations from the infinitesimal ones by means of the method of
differential invariants.

A local function $\Omega=\Omega(z^{A})$ on jet space is a differential
invariant of the group of transformations provided that
\cite{StH,OlP,SoL,Blu}
\begin{equation}\label{eq:diffinv}
\Omega(\tilde{z}^{A})=\Omega(z^{A}),
\end{equation}
i.e.\ they are jet space functions that keep their local functional
form under the action of the one-parameter group. Since the infinitesimal
action is characterized by the corresponding generator \eqref{eq:genjs}, a
differential invariant must be constant along the integral curves of the
generator
\begin{equation}\label{eq:diffeq}
\bm{X}(\Omega)=X^{A}\bm{\partial_{A}}(\Omega)=
X^{1}\frac{\partial\Omega}{\partial{z^1}}+\cdots
+X^{m}\frac{\partial\Omega}{\partial{z^m}}=0.
\end{equation}
This is a linear homogeneous first order partial differential equation which
has $m-1$ functionally independent solutions, where $m$ is the jet space dimension.
These functionally independent differential invariants are
just the $m-1$ integration constants, $\Omega_i(z^{A})=c_i$, of the
corresponding \emph{characteristic system} of ordinary differential equations
\begin{equation}\label{eq:chareq}
\frac{\mathrm{d}z^{1}}{X^{1}(z^{A})}=
\frac{\mathrm{d}z^{2}}{X^{2}(z^{A})}=\cdots=
\frac{\mathrm{d}z^{m}}{X^{m}(z^{A})}.
\end{equation}
The idea of the method is to eliminate $m-1$ coordinates from the
differential invariants and rewrite the equation for the remaining
coordinate, lets say $z^{\check{B}}$, as an autonomous first order ordinary
equation for this single variable,
\begin{equation}
\frac{\mathrm{d}z^{\check{B}}}
{X^{\check{B}}(z^{\check{B}},\Omega_1,\ldots,\Omega_{m-1})}=
\mathrm{d}\varepsilon,
\end{equation}
which is integrable at least in quadrature. Integrating this equation with
general initial conditions $z^{A}$ at $\varepsilon=0$ and denoting the
solutions for a generic value of the parameter as $\tilde{z}^{A}$, we obtain
the one-parameter transformation
$\tilde{z}^{\check{B}}=\tilde{z}^{\check{B}}(z^{A};\varepsilon)$ after
substituting all the differential invariants evaluated on the full initial
conditions. The remaining transformations are found by isolating the other
$m-1$ coordinates from the definitions of the differential invariants
$\Omega_i(\tilde{z}^{A})=\Omega_i(z^{A})$. In our opinion, this method is
more efficient and straightforward than the standard exponentiation of the
vector fields. We present some examples which make the involved
procedure more clear.

We start with the well-known example of the $\mathrm{SO}(2)$ group acting on
the plane, which has as generator
\begin{equation}\label{eq:so2gen}
\bm{X}=-y\bm{\partial_{x}}+x\bm{\partial_{y}}.
\end{equation}
In this case the characteristic equation
\begin{equation}\label{eq:chareqso2gen}
-\frac{\mathrm{d}x}{y}=\frac{\mathrm{d}y}{x},
\end{equation}
can be rewritten as
\begin{equation}
\mathrm{d}(x^{2}+y^{2})=0,
\end{equation}
giving rise to the single differential invariant
\begin{equation}\label{eq:diffinvso2}
\Omega_1(x,y)=x^{2}+y^{2}.
\end{equation}
In order to find the finite transformation for $x$ we must integrate its
ordinary equation
\begin{align}
\mathrm{d}\varepsilon&=-\frac{\mathrm{d}x}{y}\nonumber\\
&=-\frac{\mathrm{d}x}{\sqrt{\Omega_1-x^{2}}}, \label{eq:finitex}
\end{align}
making use of the differential invariant (\ref{eq:diffinvso2}), which is
constant along the integral curves of the $\mathrm{SO}(2)$ generator.
Integrating equation (\ref{eq:finitex}) from $0$ to $\varepsilon$ and
denoting the initial conditions by $(x,y)$ and the solutions by
$(\tilde{x},\tilde{y})$ yields
\begin{equation}\label{eq:tex}
\tilde{x}=\cos(\varepsilon)x-\sin(\varepsilon)y,
\end{equation}
after substituting back the differential invariant evaluated at the initial
conditions. The remaining transformation is found from the definition of the
differential invariant since
\begin{equation}
\tilde{x}^{2}+\tilde{y}^{2}=x^{2}+y^{2},
\end{equation}
then solving for $\tilde{y}$ and substituting equation (\ref{eq:tex}) gives
\begin{equation}\label{eq:tey}
\tilde{y}=\sin(\varepsilon)x+\cos(\varepsilon)y.
\end{equation}
The transformations (\ref{eq:tex}) and (\ref{eq:tey}) are the standard
rotations in the $(x,y)$ plane as expected.

The next example is the following three-dimensional generator
\begin{equation}\label{eq:vecf}
\bm{X}=-y\bm{\partial_{x}}+x\bm{\partial_{y}}
+(1+z^{2})\bm{\partial_{z}},
\end{equation}
having as characteristic system
\begin{equation}\label{eq:chareqvecf}
-\frac{\mathrm{d}x}{y}=\frac{\mathrm{d}y}{x}=\frac{\mathrm{d}z}{1+z^{2}}.
\end{equation}
In this case we have two differential invariants and the first equality
yields the same differential invariant presented in the previous example
\eqref{eq:diffinvso2}. Consequently, the planar rotations \eqref{eq:tex} and
\eqref{eq:tey} hold again. The second equality can be written as
\begin{align}
\frac{\mathrm{d}z}{1+z^{2}}&=\frac{\mathrm{d}y}{x}\nonumber\\
&=\frac{\mathrm{d}y}{\sqrt{\Omega_1-y^{2}}}, \label{eq:diffinvvecf}
\end{align}
by means of the first differential invariant, giving the exact differential
\begin{equation}
\mathrm{d}\left(\arctan{z}-\arcsin\frac{y}{\sqrt{\Omega_1}}\right)=0,
\end{equation}
whose integration gives us the second differential invariant. Since any
function of a differential invariant inherits that property as well, we
choose
\begin{equation}
\Omega_2(\vec{x})=\tan\biggl(\arctan{z}-\arcsin\frac{y}{\sqrt{\Omega_1}}
\biggr)=\frac{xz-y}{xz+x}.
\label{eq:diffinvvecf2}
\end{equation}
Using now the definition for a differential invariant we have the following
relation between the solution $\vec{\tilde{x}}$ and the initial condition
$\vec{x}$
\begin{equation}
\frac{\tilde{x}\tilde{z}-\tilde{y}}{\tilde{y}\tilde{z}+\tilde{x}}
=\frac{xz-y}{yz+x},
\end{equation}
from which after substituting the planar rotations \eqref{eq:tex} and
\eqref{eq:tey} and isolating $\tilde{z}$ we obtain the
$\mathrm{SL(2,\!I\!R)}$ transformation
\begin{equation}\label{eq:tez}
\tilde{z}=\frac{\cos(\varepsilon)z+\sin(\varepsilon)}
{-\sin(\varepsilon)z+\cos(\varepsilon)}.
\end{equation}
The transformations \eqref{eq:tex}, \eqref{eq:tey} and \eqref{eq:tez} are the
finite version of the infinitesimal ones described by the generator
\eqref{eq:vecf}.

The last example is one where the generator is characterized by an arbitrary
dependence. Those cases involve general subclasses of diffeomorphisms and
represent infinite dimensional algebras. We choose the generator
\eqref{eq:xxicol} of the Collinson \emph{Ansatz},
\begin{equation*}
\bm{\hat{X}}_{\xi^{z}}=\xi^{z}\bm{\partial_{z}}
+\overline{\xi^{z}}\bm{\partial_{\bar{z}}}
+\frac{1}{2}\left(\partial_{z}\xi^{z}
+\partial_{\bar{z}}\overline{\xi^{z}}\right)\bm{\partial_{P}},
\end{equation*}
whose characteristic system is
\begin{equation}
\frac{\mathrm{d}z}{\xi^{z}(z)}=\frac{\mathrm{d}\bar{z}}{\overline{\xi^{z}}(\bar{z})}
=\frac{2\mathrm{d}P}{\partial_{z}\xi^{z}+\partial_{\bar{z}}\overline{\xi^{z}}}.
\label{eq:chareqxx}
\end{equation}
Here $\xi^{z}(z)$ is an arbitrary holomorphic function on the complex
$z$ plane. In order for this generator to be uniquely integrable this function
must be nonvanishing, hence, this component represents an arbitrary conformal
transformation, i.e.\ a general one-parameter holomorphism
\begin{subequations}\label{eq:tozxx}
\begin{equation}
\tilde{z}=\tilde{z}(z;\varepsilon),\qquad
\xi^{z}(z)\equiv{
\frac{\partial\tilde{z}}{\partial\varepsilon}
\hspace{0.05in}\vline}_{\hspace{0.05in}\varepsilon=0},
\end{equation}
with nonvanishing derivative
\begin{equation}\label{eq:dtz}
\frac{\mathrm{d}\tilde{z}}{\mathrm{d}z}
=\frac{\tilde{\xi}^{z}(\tilde{z})}{\xi^{z}(z)}
\neq0,
\end{equation}
\end{subequations}
and which thus is holomorphically invertible. Now, equation \eqref{eq:chareqxx} can be
rewritten in a way allowing us to identify the corresponding differential
invariant,
\begin{equation}
\mathrm{d}\left(P-\ln\left|\xi^{z}\right|\right)=0 \quad\Rightarrow\quad
\Omega(z,P)=P-\ln\left|\xi^{z}\right|.
\end{equation}
By definition, using again the notation of writing the solutions with tilde
and the initial conditions without it, the differential invariant must comply
with
\begin{equation}\label{eq:z13b}
\tilde{P}-\ln|\tilde{\xi}^{z}|=P-\ln\left|\xi^{z}\right|,
\end{equation}
from which we derive the remaining transformation
\begin{equation}\label{eq:topxx}
\tilde{P}=P+\ln\left|\frac{\mathrm{d}\tilde{z}}{\mathrm{d}z}\right|.
\end{equation}
In summary, the generator \eqref{eq:xxicol} of the Collinson \emph{Ansatz}
represents the infinitesimal version of a conformal transformation
\eqref{eq:tozxx} compensated for with the supertranslation \eqref{eq:topxx} on
the structural function $P$.

The three examples studied in this appendix are representative of the
procedures needed to find the finite version of the generators appearing
along the paper, by means of the use of the differential invariant method.

\section{\label{app:Spher}Spherically symmetric \emph{Ansatz}, details}

In this appendix and the following ones we provide the details of the
calculations concerning the remaining examples we study in the paper. This
appendix is devoted to the spherically symmetric \emph{Ansatz}. We start with the
static case \eqref{eq:ssa} having jet space coordinates
$z^{A}=\left(t,r,\theta,\varphi,F,N,Y\right)$. Since all the structural
functions are radial functions then, according to the classification of
Sec.~\ref{sec:Crit}, $u^{\bar{I}}=u^{I}=(F,N,Y)$,
$x^{\bar{\alpha}}=\left(r\right)$ and
$x^{\hat{\alpha}}=\left(t,\theta,\varphi\right)$, which reduce the
complementary conditions \eqref{eq:critxi} to
\begin{equation}\label{eq:crits}
\partial_{t}\eta^{I}=\partial_{\theta}\eta^{I}=\partial_{\varphi}\eta^{I}
=\partial_{t}\xi^{r}=\partial_{\theta}\xi^{r}=\partial_{\varphi}\xi^{r}=0.
\end{equation}
Therefore, the generator of infinitesimal residual symmetries must have the
following general form
\begin{align}
\bm{X}={}&\xi^{t}(t,r,\theta,\varphi)\bm{\partial_t}
+\xi^{r}(r)\bm{\partial_r}
+\xi^{\theta}(t,r,\theta,\varphi)\bm{\partial_\theta}\nonumber\\
&+\xi^{\varphi}(t,r,\theta,\varphi)\bm{\partial_\varphi}
+\eta^{F}(r,F,N,Y)\bm{\partial_F}\nonumber\\
&+\eta^{N}(r,F,N,Y)\bm{\partial_N}
+\eta^{Y}(r,F,N,Y)\bm{\partial_Y}.\label{eq:gensss}
\end{align}

The spherical Lie-derivative criterion \eqref{eq:crita} reads
\begin{subequations}\label{eq:critasss}
\begin{align}
N\eta^{F}+2F\eta^{N}+2NF\partial_{t}\xi^{t}&=0,\label{c1}\\
\eta^{F}-2F\partial_{r}\xi^{r}&=0,\label{c2}\\
\eta^{Y}+Y\partial_{\theta}\xi^{\theta}&=0,\label{c3}\\
\eta^{Y}+Y\left(\partial_{\varphi}\xi^{\varphi}
+\cot(\theta)\xi^{\theta}\right)&=0,\label{c4}\\
\partial_{r}\xi^{t}=\partial_{r}\xi^{\theta}=
\partial_{r}\xi^{\varphi}&=0,\label{c5}\\
-N^{2}F\partial_{\theta}\xi^{t}+Y^{2}\partial_{t}\xi^{\theta}&=0,\label{c6}\\
-N^{2}F\partial_{\varphi}\xi^{t}
+Y^{2}\sin^{2}(\theta)\partial_{t}\xi^{\varphi}&=0,\label{c7}\\
\partial_{\varphi}\xi^{\theta}
+\sin^{2}(\theta)\partial_{\theta}\xi^{\varphi}&=0.\label{c10}
\end{align}
\end{subequations}

From Eq.~\eqref{c5} and using in Eqs.~(\ref{c6})-(\ref{c7}) the fact that the
structural functions of the \emph{Ansatz} are free variables from which the
generator components along spacetime are independent, we conclude that
$\xi^{t}$ is a function of $t$ only and $\xi^{\theta}$, $\xi^{\varphi}$ are
functions of $\theta$ and $\varphi$ only. Using this in equations \eqref{c1},
\eqref{c3} and \eqref{c4} and that the generator components along the
structural functions are independent of $t$, $\theta$ and $\varphi$, these
equations are only satisfied if
\begin{subequations}\label{eq:sysxis}
\begin{align}
\partial_{t}\xi^{t}&=C_1,\label{c11}\\
\partial_{\theta}\xi^{\theta}&=C_2
=\partial_{\varphi}\xi^{\varphi}+\cot\theta\xi^{\theta},\label{c12}
\end{align}
\end{subequations}
where $C_{1}$ and $C_{2}$ are constants. This subsystem can be solved
together with Eq.~\eqref{c10}. In fact, deriving Eq.~\eqref{c10} with respect
to $\varphi$ and using Eq.~\eqref{c12} we obtain that $C_2=0$ and
$\partial_{\varphi}^2\xi^{\theta}=-\xi^{\theta}$, from which the integration
is straightforward
\begin{subequations}\label{eq:solxis}
\begin{align}
\xi^{t}&=C_{1}t+\kappa_{4},\label{eq:xitsss}\\
\xi^{\theta}&=\kappa_1\sin\varphi+\kappa_2\cos\varphi,\label{eq:xithsss}\\
\xi^{\varphi}&=(\kappa_1\cos\varphi - \kappa_2\sin\varphi)\cot\theta+\kappa_{3},
\label{eq:xipsss}
\end{align}
\end{subequations}
and the $\kappa$'s are new integration constants. Now it only remains to solve
the consistent algebraic system \eqref{c1}-\eqref{c3} for the generator
components along the structural functions, which results after substituting
the solutions \eqref{eq:solxis}, that gives us
\begin{subequations}\label{eq:soletas}
\begin{align}
\eta^{N}&=-\left(\partial_r\xi^{r}+C_{1}\right)N,\label{eq:etanssa}\\
\eta^{F}&=2\partial_r\xi^{r}F,\label{eq:etafssa}\\
\eta^{Y}&=0\label{eq:etayssa}.
\end{align}
\end{subequations}
This allows us to arrive to the final form of the generator (\ref{eq:gensss}) of
residual symmetries on static spherically symmetric spacetimes
\begin{align}
\bm{X}={}&C_{1}(t\bm{\partial_t}-N\bm{\partial_N})\nonumber\\
&+\xi^{r}(r)\bm{\partial_r}
+\partial_r\xi^{r}\left(2F\bm{\partial_F}-N\bm{\partial_N}\right)\nonumber\\
&+\kappa_1(\sin\varphi\bm{\partial_\theta}
+\cot\theta \cos\varphi\bm{\partial_\varphi})\nonumber\\
&+\kappa_2(\cos\varphi\bm{\partial_\theta}
-\cot\theta \sin\varphi\bm{\partial_\varphi})\nonumber\\
&+\kappa_{3}\bm{\partial_\varphi}+\kappa_{4}\bm{\partial_t},
\label{eq:genssa}
\end{align}
which is a linear combination of the generators
\eqref{eq:x1ssa}-\eqref{eq:kil3ssa} exhibit in the main text.

We continue by studying the differences with respect to the previous behavior
of considering a time-dependent spherically symmetric \emph{Ansatz} \eqref{eq:ss}.
Incorporating time dependency changes the jet space classification of
Sec.~\ref{sec:Crit} to $u^{\bar{I}}=u^{I}=(F,N,Y)$,
$x^{\bar{\alpha}}=\left(t,r\right)$ and
$x^{\hat{\alpha}}=\left(\theta,\varphi\right)$, giving now the complementary
conditions
\begin{equation}
\partial_{\theta}\eta^{I}=\partial_{\varphi}\eta^{I}
=\partial_{\theta}\xi^{t}=\partial_{\varphi}\xi^{t}
=\partial_{\theta}\xi^{r}=\partial_{\varphi}\xi^{r}=0,\label{eq:critts}
\end{equation}
which modify the generator \eqref{eq:gensss} accordingly. In the residual
criterion \eqref{eq:critasss}, the equations \eqref{c1}-\eqref{c4} which
algebraically determine the generator components along the structural
functions remain unchanged, but the rest is transformed to
\begin{subequations}
\begin{align}
\partial_{t}\xi^{r}-N^2F^2\partial_{r}\xi^{t}&=0,\label{ct5}\\
\partial_{t}\xi^{\theta}=\partial_{r}\xi^{\theta}
=\partial_{t}\xi^{\varphi}=\partial_{r}\xi^{\varphi}&=0,\label{ct6}\\
\partial_{\varphi}\xi^{\theta}
+\sin^{2}(\theta)\partial_{\theta}\xi^{\varphi}&=0.\label{cat10}
\end{align}
\end{subequations}
Performing a similar analysis as in the static case we end with the same
solution except for the component $\xi^{t}$, which is now an arbitrary
function of $t$, and this modify the component along $N$ as follows
\begin{equation}
\eta^{N}=-\left(\partial_{r}\xi^{r}+\partial_{t}\xi^{t}\right)N.
\label{eq:etanss}
\end{equation}
Consequently, in the time-dependent case the generator \eqref{eq:genssa} then
changes to
\begin{align}
\bm{X}={}&\xi^{t}(t)\bm{\partial_t}
-\partial_{t}\xi^{t}N\bm{\partial_N}\nonumber\\
&+\xi^{r}(r)\bm{\partial_r}
-\partial_{r}\xi^{r}\left(N\bm{\partial_N}-2F\bm{\partial_F}\right)\nonumber\\
&+\kappa_1(\sin\varphi\bm{\partial_\theta}
+\cot\theta \cos\varphi\bm{\partial_\varphi})\nonumber\\
&+\kappa_2(\cos\varphi\bm{\partial_\theta}
-\cot\theta \sin\varphi\bm{\partial_\varphi})+\kappa_{3}\bm{\partial_\varphi}.
\label{eq:genss}
\end{align}

\section{\label{app:AdS}AdS waves, details}

In the jet space coordinates for the $D$-dimensional AdS waves
\eqref{eq:AdSw}, $z^{A}=\left(u,v,y,\vec{x},F\right)$, the structural
function $u^{\bar{I}}=u^{I}=(F)$ depends only on the retarded time and the
front-wave coordinates $x^{\bar{\alpha}}=\left(u,y,\vec{x}\right)$, i.e.\ it
is independent of the null rays parameter $x^{\hat{\alpha}}=(v)$. Hence, the
complementary conditions \eqref{eq:critxi} take the form
\begin{equation}\label{eq:ccritads}
\partial_{v}\eta^{F}=\partial_{v}\xi^{u}=
\partial_{v}\xi^{y}=\partial_{v}\xi^{i}=0,
\end{equation}
giving the following general form for the generator of infinitesimal residual
symmetries
\begin{align}
\bm{X}={}&\xi^{u}(u,y,\vec{x})\bm{\partial_u}
+\xi^{v}(u,v,y,\vec{x})\bm{\partial_v}
+\xi^{y}(u,y,\vec{x})\bm{\partial_{y}}
\nonumber\\
&+\bm{\vec{\xi}}(u,y,\vec{x})+\eta^{F}(u,y,\vec{x},F)\bm{\partial_F},
\label{eq:genadsw}
\end{align}
where $i=1,\ldots,D-3$ and we use the standard notation for Euclidean vectors
$\bm{\vec{v}}=v^{i}\bm{\partial_{i}}$.

The Lie-derivative criterion \eqref{eq:crita} gives the following system of
equations
\begin{subequations}
\begin{align}
\eta^{F}+2\left(\partial_{u}\xi^{u}-\frac{\xi^{y}}{y}\right)F
+2\partial_{u}\xi^{v}&=0,\label{e11}\\
\partial_{u}\xi^{u}+\partial_{v}\xi^{v}-\frac{2}{y}\xi^{y}&=0,\label{e12}\\
\partial_{y}\xi^{u}=\partial_{i}\xi^{u}&=0,\frac{}{}\label{e15}\\
\partial_{y}\xi^{v}-\partial_{u}\xi^{y}&=0,\frac{}{}\label{e13}\\
\partial_{i}\xi^{v}-\partial_{u}\xi_{i}&=0,\frac{}{}\label{e14}\\
\partial_{y}\xi^{y}-\frac{\xi^{y}}{y}&=0,\label{e17}\\
\partial_{i}\xi^{y}+\partial_{y}\xi_{i}&=0,\frac{}{}\label{e18}\\
\partial_{(i}\xi_{j)}-\frac{\xi^{y}}{y}\delta_{ij}&=0,\label{e19}
\end{align}
\end{subequations}
where we denote $\xi_i=\delta_{ij}\xi^j$. The first obvious conclusion from
\eqref{e15} is that the component along the retarded time only depend of the
retarded time, $\xi^{u}=\xi^{u}(u)$. Second, deriving the system with respect
to the null ray parameter $v$ and using the complementary conditions
\eqref{eq:ccritads} we obtain
\begin{equation}\label{e23}
\partial^{2}_{vu}\xi^{v}=\partial^{2}_{vv}\xi^{v}=
\partial^{2}_{vy}\xi^{v}=\partial^{2}_{vi}\xi^{v}=0,
\end{equation}
which means the dependence of $\xi^v$ in $v$ is linear and separable in sum
with respect to the rest of the coordinates
\begin{equation}\label{e24}
\xi^{v}=2C_{1}v+\mathcal{V}(u,y,\vec{x}).
\end{equation}
Then, isolating from equation \eqref{e12} we obtain
\begin{equation}\label{e25}
\xi^{y}=\frac{y}{2}\left(\dot\xi^{u}+2C_{1}\right),
\end{equation}
where the dot represents derivative with respect to the retarded time $u$.
This implies that $\xi^{y}$ is function of $u$ and $y$ only, using this fact
in Eq.~(\ref{e18}) to allow us to conclude that the spatial components $\xi_{i}$
are functions of $u$ and $\vec{x}$ only. Additionally, the symmetric part of
their spatial derivatives $\partial_{i}\xi_{j}$ is obtained from
Eq.~\eqref{e19} as
\begin{equation}\label{eq:symxii}
\partial_{(i}\xi_{j)}=\frac{1}{2}(\dot{\xi}^{u}+2C_1)\delta_{ij},
\end{equation}
and consequently, the derivatives themselves are given by
\begin{equation}\label{eq:d_i_xi_j}
\partial_{i}\xi_{j}=\frac{1}{2}(\dot{\xi}^{u}+2C_1)\delta_{ij}
+\beta_{ij}(u,\vec{x}),
\end{equation}
where $\beta_{ij}=-\beta_{ji}$ is the antisymmetric part of the derivatives.
Now deriving the symmetric part \eqref{eq:symxii} with respect to $x^{l}$ we
end with
\begin{subequations}
\begin{equation}
\partial^2_{il}\xi_{j}+\partial^2_{lj}\xi_{i}=0,\label{eq:perm0}
\end{equation}
and permuting indices we can also write
\begin{align}
\partial^2_{lj}\xi_{i}+\partial^2_{ji}\xi_{l}&=0,\label{eq:perm1}\\
\partial^2_{ji}\xi_{l}+\partial^2_{il}\xi_{j}&=0.\label{eq:perm2}
\end{align}
\end{subequations}
Adding Eqs.~\eqref{eq:perm0} and \eqref{eq:perm2} and subtracting
Eq.~\eqref{eq:perm1} we arrive at
\begin{equation}
2\partial_{l}\partial_{i}\xi_{j}=2\partial_{l}\beta_{ij}=0.
\end{equation}
Then, the $\beta_{ij}$'s are functions of $u$ only. This allows to
straightforwardly integrate \eqref{eq:d_i_xi_j} as
\begin{align}
\xi_{j}&=\frac{1}{2}(\dot{\xi}^{u}+2C_1)x_{j}+\beta_{ij}x^{i}+P_{j}(u).
\label{eq:xij}
\end{align}
Using all this information Eqs.~\eqref{e13} and \eqref{e14} reduce to
\begin{subequations}\label{eq:sysV}
\begin{align}
\partial_{y}\mathcal{V}&=\frac{y}{2}\ddot{\xi}^{u},\label{eq:dfdy}\\
\partial_{i}\mathcal{V}&=\frac{x_{i}}{2}\ddot{\xi}^{u}+\dot{\beta}_{ji}x^{j}
+\dot{P}_{i}\label{eq:dfdi}.
\end{align}
\end{subequations}
Taking the integrability conditions of Eq.~\eqref{eq:dfdi} we get
\[
0=\partial_{[ji]}\mathcal{V}=\dot{\beta}_{ji},
\]
concluding that the antisymmetric matrix is in fact a constant one,
$\beta_{ij}=C_{ij}$. The system \eqref{eq:sysV} is then integrable and from
\eqref{e24} the general solution for the component along the null ray is
given by
\begin{equation}
\xi^{v}=2C_{1}v+\frac{1}{4}(y^{2}+\vec{x}^2)\ddot\xi^{u}
+\dot{\vec{P}}\cdot\vec{x}+\alpha(u).\label{e38}
\end{equation}
Finally, we determine the component along the structural function from
Eq.~\eqref{e11}
\begin{equation}
\eta^{F}=2C_{1}F-\dot\xi^{u}
-\frac{1}{2}\left(y^2+\vec{x}^2\right)\dddot{\xi^{u}}
-2\ddot{\vec{P}}\cdot\vec{x}-2\dot{\alpha}.
\label{eq:etaf}
\end{equation}
When we substitute equations \eqref{e25}, \eqref{eq:xij}, \eqref{e38} and
\eqref{eq:etaf} in the generator \eqref{eq:genadsw} we obtain
\begin{align}
\bm{X}={}&C_{1}\left(2v\bm{\partial_v}+y\bm{\partial_y}
+x^{i}\bm{\partial_{i}}+2F\bm{\partial_F}\right)\frac{}{}\nonumber\\
&+\frac12{C}^{ij}\left(x_{i}\bm{\partial}_{\bm{j}}
-x_{j}\bm{\partial}_{\bm{i}}\right)\nonumber\\
&+\alpha(u)\bm{\partial_v}-2\dot\alpha\bm{\partial_F}\frac{}{}\nonumber\\
&+\xi^{u}(u)\bm{\partial_u}+\frac{\dot\xi^{u}}{2}\left(y\bm{\partial_y}
+x^{i}\bm{\partial_{i}}-2F\bm{\partial_F}\right)\nonumber\\
&+\frac{1}{4}\left(y^2+\vec{x}^2\right)
\left(\ddot\xi^{u}\bm{\partial_v}
-2\dddot{\xi^{u}}\bm{\partial_F}\right)\nonumber\\
&+\bm{\vec{P}}(u)+\vec{x}\cdot\!\left(\dot{\vec{P}}\bm{\partial_{v}}
-2\ddot{\vec{P}}\bm{\partial_{F}}\right),\frac{}{}
\label{eq:genads}
\end{align}
where the spatial indices $(i,j)$ are raised and lowered using the Euclidean
metric $\delta_{ij}$.

\subsection*{Finite reparameterization of the retarded time}

The result of the previous derivation is a superposition of the vector fields
\eqref{eq:x1ads}-\eqref{eq:kil2ads} we report in Sec.~\ref{sec:AdS}. Using
the method of differential invariants explained in App.~\ref{app:inv}, it is
very easy to integrate most of these vector fields to show that the
corresponding finite transformations are those also reported in the same
section. The integration of the generator \eqref{eq:xuads} is a little more
difficult since it comprise an arbitrary function similarly to the example
related with the Collinson \emph{Ansatz} we study in App.~\ref{app:inv}. Since the
present case is more involved we found convenient to give the details of its
integration. The characteristic system for the generator (\ref{eq:xuads}) is
\begin{align}
\frac{\mathrm{d}u}{\xi^{u}}=\frac{2\mathrm{d}y}{\dot{\xi}^{u}y}
=\frac{2\mathrm{d}x^1}{\dot{\xi}^{u}x^1}=\cdots
&=\frac{2\mathrm{d}x^{D-3}}{\dot{\xi}^{u}x^{D-3}}\nonumber\\
&=\frac{4\mathrm{d}v}{\ddot{\xi}^{u}(y^2+\vec{x}^2)}\nonumber\\
&=\frac{-2\mathrm{d}F}{2\dot{\xi}^{u}F+\dddot{\xi^{u}}(y^2+\vec{x}^2)},
\end{align}
and is equivalent to the set of equations
\begin{subequations}
\begin{align}
\frac{\mathrm{d}x^{\hat{\imath}}}
{x^{\hat{\imath}}}&=\frac{\dot\xi^{u}\mathrm{d}u}{2\xi^{u}},
\label{eq:cs_uxi}\\
\mathrm{d}v&=\frac14\frac{x_{\hat{\imath}}x^{\hat{\imath}}}{\xi^{u}}
\ddot\xi^{u}\mathrm{d}u,\label{eq:cs_uv}\\
\xi^{u}\mathrm{d}F&=-\left(\dot\xi^{u}F
+\frac{1}{2}\dddot{\xi^{u}}x_{\hat{\imath}}x^{\hat{\imath}}\right)
\mathrm{d}u,\label{eq:cs_uF}
\end{align}
\end{subequations}
where $x^{\hat{\imath}}=(y,x^{1},...,x^{D-3})$ are the wave-front coordinates
and no sum is understood in the first equation. We proceed as we did for the
Collinson \emph{Ansatz}, remembering that the arbitrary function
$\xi^{u}=\xi^{u}(u)$ is the infinitesimal version of a general
reparameterization of the retarded time $\tilde{u}=\tilde{u}(u;\varepsilon)$
and
\begin{equation}
\frac{\mathrm{d}\tilde{u}}{\mathrm{d}u}=
\frac{\tilde{\xi}^{u}(\tilde{u})}{\xi^{u}(u)}.
\label{eq:e51}
\end{equation}
This relation will allow us to eliminate the appearances of the infinitesimal
contributions of the reparameterization in favor of the finite one and their
derivatives. We begin with the first characteristic equations
\eqref{eq:cs_uxi} which can be straightforwardly integrated given the first
$D-2$ differential invariants,
\begin{equation}\label{eq:e52}
\mathrm{d}\ln\left(\frac{x^{\hat{\imath}}}{\sqrt{\xi^{u}}}\right)=0,
\quad \Rightarrow \quad
\Omega^{\hat{\imath}}(u,x^{\hat{\imath}})=
\frac{x^{\hat{\imath}}}{\sqrt{\xi^{u}}}.
\end{equation}
Using their invariant property
\begin{equation}
\frac{\tilde{x}^{\hat{\imath}}}{\sqrt{\tilde{\xi}^{u}(\tilde{u})}}
=\frac{x^{\hat{\imath}}}{\sqrt{\xi^{u}(u)}},
\end{equation}
and the relation \eqref{eq:e51} we find the $D-2$ finite transformations for
the wave-front coordinates
\begin{equation}\label{e53}
\tilde{x}^{\hat{\imath}}=
\sqrt{\frac{\mathrm{d}\tilde{u}}{\mathrm{d}u}}x^{\hat{\imath}}.
\end{equation}
For solving the second characteristic equation \eqref{eq:cs_uv} we use that
any function of a differential invariant is also a differential invariant,
hence, from definition \eqref{eq:e52} we get that
\begin{equation}\label{eq:e54}
\Omega_{\hat{\imath}}\Omega^{\hat{\imath}}=
\frac{x_{\hat{\imath}}x^{\hat{\imath}}}{\xi^{u}},
\end{equation}
is a differential invariant. Replacing it in Eq.~\eqref{eq:cs_uv} we arrive
to a new differential invariant
\begin{equation}\label{eq:e55}
\mathrm{d}\left(v
-\frac14\Omega_{\hat{\imath}}\Omega^{\hat{\imath}}\dot\xi^{u}\right)=0,
\quad \Rightarrow \quad
\Omega_v(u,v)=v-\frac14\Omega_{\hat{\imath}}\Omega^{\hat{\imath}}\dot\xi^{u}.
\end{equation}
Using their invariant property $\Omega_v(\tilde{u},\tilde{v})=\Omega_v(u,v)$
and the derivative of the relation \eqref{eq:e51} we obtain the finite
transformation for the null ray parameter
\begin{align}
\tilde{v}&=v+\frac{1}{4}x_{\hat{\imath}}x^{\hat{\imath}}
\frac{\dot\xi^{u}(\tilde{u})-\dot\xi^{u}(u)}{\xi^{u}(u)}\nonumber\\
&=v+\frac{1}{4}x^{\hat{\imath}}x_{\hat{\imath}}
\frac{\mathrm{d}}{\mathrm{d}u}
\ln\left(\frac{\mathrm{d}\tilde{u}}{\mathrm{d}u}\right).
\label{e58}
\end{align}
The last characteristic equation \eqref{eq:cs_uF} can be rewritten using
\eqref{eq:e54} as
\begin{equation}
\xi^{u}\mathrm{d}F+\left(\dot\xi^{u}F
+\frac{1}{2}\Omega_{\hat{\imath}}\Omega^{\hat{\imath}}\xi^{u}
\dddot\xi^{u}\right)du=0,
\label{e59}
\end{equation}
which is an integrable equation since the crossed partial derivatives
coincide. Hence, the equation is an exact differential of the last
differential invariant,
\begin{equation}\label{e65}
\Omega_F(u,F)=\xi^{u}F
+\frac{\Omega_{\hat{\imath}}\Omega^{\hat{\imath}}}{2}
\left(\xi^{u}\ddot\xi^{u}-\frac{1}{2}(\dot\xi^{u})^{2}\right).
\end{equation}
From their invariance $\Omega_F(\tilde{u},\tilde{F})=\Omega_F(u,F)$ and using
the second derivative of the relation \eqref{eq:e51} we get the finite
transformation for the structural function
\begin{equation}\label{e69b}
\tilde{F}=\left(\frac{\mathrm{d}\tilde{u}}{\mathrm{d}u}\right)^{-1}
\left[F+\left(\frac{\mathrm{d}\tilde{u}}{\mathrm{d}u}\right)^{1/2}
\frac{\mathrm{d}^2}{\mathrm{d}u^2}
\left(\frac{\mathrm{d}\tilde{u}}{\mathrm{d}u}\right)^{-1/2}
x_{\hat{\imath}}x^{\hat{\imath}}\right].
\end{equation}
Eqs.~(\ref{e53}), (\ref{e58}) and (\ref{e69b}) are the way in which the rest
of the variables involved in the AdS wave \emph{Ansatz} \eqref{eq:AdSw} must change
in order to compensate an arbitrary reparameterization of the retarded time
$\tilde{u}=\tilde{u}(u)$, this is precisely the transformation
\eqref{eq:t3ads} reported in the main text.

\subsection*{AdS from AdS waves}

The last issue we address in this appendix is the derivation of the most
general way to locally express the AdS spacetime with the AdS wave \emph{Ansatz}
\eqref{eq:AdSw}, which gives just the structural function \eqref{eq:AdS}.
This is relevant to the interpretation of the infinite-dimensional subgroup
of the connected group of residual symmetries we provide at the end of
Sec.~\ref{sec:AdS}. AdS is defined as the spacetime having constant negative
curvature in any dimension $D$ and we need to impose this condition on the
AdS waves. Spacetime curvature is in general decomposed in their completely
traceless part represented in the Weyl tensor, $W^\alpha{}_{\beta\mu\nu}$,
and their first trace defining the Ricci tensor, $R_{\alpha\beta}$. In turn,
the Ricci tensor has a traceless part,
$S_{\alpha\beta}=R_{\alpha\beta}-Rg_{\alpha\beta}/D$, and its trace which is
the scalar curvature, $R$. Hence, the Weyl tensor, the traceless part of the
Ricci tensor and the scalar curvature completely determine the full
curvature. Spaces where one or both of the traceless parts vanish have
special properties: due to the conformal invariance of the Weyl tensor its
vanishing defines the conformally flat spacetimes, a vanishing traceless part
of the Ricci tensor defines the so-called Einstein spaces and when both
tensors vanish the full curvature is completely determined by the scalar
curvature which is necessarily constant. The latter are the constant
curvature spaces. Since the AdS waves \eqref{eq:AdSw} already have constant
negative scalar curvature, we only impose they are conformally flat and at
the same time Einstein spaces. The traceless curvature tensors for these
backgrounds are given by
\begin{align}
W_{\alpha\beta\mu\nu}&=\frac{2l^2}{y^2}
\left(\partial^2_{\hat{\imath}\hat{\jmath}}F
-\frac{\partial^{\hat{k}}\partial_{\hat{k}}F}{D-2}
\delta_{\hat{\imath}\hat{\jmath}}\right)
\delta_{[\alpha}^u\delta_{\beta]}^{\hat{\imath}}
\delta_{[\mu}^u\delta_{\nu]}^{\hat{\jmath}},\label{eq:WeylAdSw}\\
S_{\alpha\beta}&=\frac12
\left(\partial^{\hat{k}}\partial_{\hat{k}}F
-\frac{D-2}{y}\partial_yF\right)
\delta_{\alpha}^u\delta_{\beta}^u.\label{eq:tlRcAdSw}
\end{align}
The vanishing of the Weyl components
\begin{equation}
W_{u\hat{\imath}u\hat{\jmath}}=\frac{l^2}{2y^2}
\partial^2_{\hat{\imath}\hat{\jmath}}F=0, \qquad
\hat{\imath}\neq\hat{\jmath},
\end{equation}
implies the structural function characterizing the profile of the AdS wave is
separable in sum in all the front-wave coordinates,
\begin{equation}\label{eq:sep}
F(u,x^{\hat{\imath}})=Y(u,y)+X^1(u,x^1)+\cdots+X^{D-3}(u,x^{D-3}).
\end{equation}
From the differences of the Weyl components
\begin{equation}
W_{uiui}-W_{uyuy}=\frac{l^2}{2y^2}\left(
\partial^2_{ii}F-\partial^2_{yy}F\right)=0,
\end{equation}
for each $i=1,\ldots,D-3$ (no sum in $i$ is understood), and using the
separability \eqref{eq:sep} we obtain
\begin{equation}\label{eq:F2}
\partial^2_{yy}Y=
\partial^2_{11}X^{1}=\cdots=
\partial^2_{{D-3},{D-3}}X^{D-3}=2F_2(u),
\end{equation}
i.e.\ the equality between the different dependencies is only possible if
there is no dependence at all on the front-wave coordinates for these second
derivatives, which in turn defines the function $F_2$ depending only on the
retarded time. Equations \eqref{eq:F2} can be easily integrated to give the
most general conformally flat AdS wave profile,
\begin{equation}\label{eq:CF}
F_\text{CF}=F_2(u)x_{\hat{\imath}}x^{\hat{\imath}}
+{F_1}_{\hat{\imath}}(u)x^{\hat{\imath}}+F_0(u).
\end{equation}
Imposing now that the last backgrounds are additionally Einstein spaces,
\begin{equation}
S_{\alpha\beta}=-\frac{D-2}2\frac{{F_1}_y}{y}
\delta_{\alpha}^u\delta_{\beta}^u=0,
\end{equation}
implies the vanishing of the function ${F_1}_y$, which reduces the conformally
flat profiles \eqref{eq:CF} to the constant curvature ones defining AdS
spacetime \eqref{eq:AdS}.

\section{\label{app:Papa}Papapetrou \emph{Ansatz}, details}

The jet space coordinates for the Papapetrou \emph{Ansatz} are
$z^{A}=\left(t,\varphi,\rho,z,X,A,h\right)$, where the structural functions
$u^{\bar{I}}=u^{I}=(X,A,h)$ are functions of the spatial coordinates
$x^{\bar{\alpha}}=(\rho,z)$ and independent of the Killing coordinates
$x^{\hat{\alpha}}=(t,\varphi)$. In this case the complementary conditions
\eqref{eq:critxi} are
\begin{equation}\label{eq:critcppp}
\partial_{t}\eta^{I}=\partial_{\varphi}\eta^{I}=
\partial_{t}\xi^{\rho}=\partial_{t}\xi^{z}=
\partial_{\varphi}\xi^{\rho}=\partial_{\varphi}\xi^{z}=0,
\end{equation}
and the generator of residual symmetries has the following general form
\begin{align}
\bm{X}={}&\xi^{t}(t,\varphi,\rho,z)\bm{\partial_t}
+\xi^{\varphi}(t,\varphi,\rho,z)\bm{\partial_\varphi}
+\xi^{\rho}(\rho,z)\bm{\partial_\rho}\nonumber\\
&+\xi^{z}(\rho,z)\bm{\partial_z}+\eta^{I}(\rho,z,u^{J})\bm{\partial}_{I}.
\label{eq:genpapapetrou}
\end{align}

The criterion (\ref{eq:crita}) in this case gives the following system of
equations for the components of the generator
\begin{subequations}
\begin{align}
\left(A^{2}+{\rho^{2}}/{X^{2}}\right)\eta^{X}+2AX\eta^{A}
+2AX\partial_{t}\xi^{\varphi}&\nonumber\\
{}+2X\left(A^{2}-{\rho^{2}}/{X^{2}}\right)\partial_{t}\xi^{t}
-\frac{2\rho}{X}\xi^{\rho}&=0,\label{d1}\\
A\eta^{X}+X\eta^{A}+\left(A^{2}-{\rho^{2}}/{X^{2}}\right)X
\partial_{\varphi}\xi^{t}&\nonumber\\
{}+X\partial_{t}\xi^{\varphi}+AX\left(\partial_{t}\xi^{t}
+\partial_{\varphi}\xi^{\varphi}\right)&=0,\label{d2}\\
\eta^{X}+2AX\partial_{\varphi}\xi^{t}
+2X\partial_{\varphi}\xi^{\varphi}&=0,\label{d5}\\
\eta^{X}-2X\eta^{h}-2X\partial_{\rho}\xi^{\rho}&=0,\label{d3}\\
\eta^{X}-2X\eta^{h}-2X\partial_{z}\xi^{z}&=0,\label{d4}\\
\partial_{\rho}\xi^{z}+\partial_{z}\xi^{\rho}&=0,\label{d6}\\
\left(A^{2}-{\rho^{2}}/{X^{2}}\right)\partial_{\rho}\xi^{t}
+A\partial_{\rho}\xi^{\varphi}&=0,\label{d7}\\
\left(A^{2}-{\rho^{2}}/{X^{2}}\right)\partial_{z}\xi^{t}
+A\partial_{z}\xi^{\varphi}&=0,\label{d8}\\
\partial_{\rho}\xi^{\varphi}+A\partial_{\rho}\xi^{t}=
\partial_{z}\xi^{\varphi}+A\partial_{z}\xi^{t}&=0.\label{d9}
\end{align}
\end{subequations}

Since the spacetime components of the generator are independent of the
structural functions, Eqs.~(\ref{d7})-(\ref{d9}) only hold if $\xi^{t}$ and
$\xi^{\varphi}$ exclusively depend on $t$ and $\varphi$. In Eqs.~\eqref{d2}
and \eqref{d5} we have dependencies on $t$ and $\varphi$ through the
derivatives of the previous components, and they are compatible with the
independence of the $\eta^I$ on $t$ and $\varphi$ \eqref{eq:critcppp}, only
if additionally
\begin{equation}
\left.\begin{aligned}
\partial_{t}\xi^{t}&=\hat{C}_1, &
\partial_{\varphi}\xi^{t}&=C_{3},\nonumber\\
\partial_{\varphi}\xi^{\varphi}&=\hat{C}_2, &
\partial_{t}\xi^{\varphi}&=C_{4},
\end{aligned}\right\}
\Longrightarrow
\left\{\begin{aligned}
\xi^{t}&=\hat{C}_{1}t+C_{3}\varphi+\kappa_{1},\\
\xi^{\varphi}&=\hat{C}_{2}\varphi+C_{4}t+\kappa_{2}.
\end{aligned}\right.
\label{eq:xitph}
\end{equation}
This makes Eqs.~\eqref{d2} and \eqref{d5} a compatible linear system for
$\eta^{X}$ and $\eta^{A}$. Inserting their solution in Eq.~\eqref{d1} it is
possible to solve also for $\xi^{\rho}$ giving
\begin{align}
\xi^{\rho}&=-(\hat{C}_{1}+\hat{C}_{2})\rho,\label{eq:xirppp}\\
\eta^{X}&=-2(\hat{C}_{2}+C_{3}A)X,\label{eq:etaxppp}\\
\eta^{A}&=-C_{4}-(\hat{C}_{1}-\hat{C}_{2})A
+C_{3}\left(A^{2}+\frac{\rho^{2}}{X^{2}}\right).\label{eq:etaappp}
\end{align}
Since $\xi^{\rho}$ is independent of $z$ it follows from equation \eqref{d6}
that $\xi^{z}$ is independent of $\rho$. Moreover, from equations \eqref{d3}
and \eqref{d4} we obtain
\begin{equation}\label{eq:xizppp}
\xi^{z}=-(\hat{C}_{1}+\hat{C}_{2})z+C_5.
\end{equation}
Finally, from Eq.~\eqref{d3} we find
\begin{equation}\label{eq:etahppp}
\eta^{h}=\hat{C}_{1}-C_{3}A.
\end{equation}
The generator \eqref{eq:genpapapetrou} is fixed as
\begin{align}
\bm{X}={}&C_{1}\left(t\bm{\partial_t}+\varphi\bm{\partial_\varphi}
-2\rho\bm{\partial_\rho}-2z\bm{\partial_z}-2X\bm{\partial_X}
+\bm{\partial_h}\right)\nonumber\\
&+C_{2}\left( t\bm{\partial_t}-\varphi\bm{\partial_\varphi}
-2A\bm{\partial_A}+2X\bm{\partial_X}+\bm{\partial_h}\right)\nonumber\\
&+C_{3}\left[\varphi\bm{\partial_t}-2AX\bm{\partial_X}
+\left(A^{2}+{\rho^{2}}/{X^{2}}\right)\bm{\partial_A}-A\bm{\partial_h}\right]
\nonumber\\
&+C_{4}(t\bm{\partial_\varphi}-\bm{\partial_A})+C_5\bm{\partial_z}
+\kappa_{1}\bm{\partial_t}+\kappa_{2}\bm{\partial_\varphi},
\label{eq:genppp}
\end{align}
where we redefined $\hat{C}_1=C_{1}+C_{2}$ and $\hat{C}_2=C_{1}-C_{2}$. This
is a linear combination of the generators \eqref{eq:gensppp} given in the
main text.

\section{\label{app:noncircColl}Noncircular Collinson \emph{Ansatz}, details}

The jet space coordinates of the noncircular Collinson \emph{Ansatz} are
extended to include the noncircular contributions as
$z^{A}=(\tau,\sigma,x,y,a,b,P,Q,M,N)$. Since all the structural functions
$u^{\bar{I}}=u^{I}=(a,b,P,Q,M,N)$ are independent of the Killing coordinates
$x^{\hat{\alpha}}=(\tau,\sigma)$ and depend only on the spatial coordinates
$x^{\bar{\alpha}}=(x,y)$, the complementary conditions \eqref{eq:critxi} are
\begin{equation}\label{eq:cccolgen}
\partial_{\tau}\eta^{I}=\partial_{\sigma}\eta^{I}=\partial_{\tau}\xi^{x}=
\partial_{\tau}\xi^{y}=\partial_{\sigma}\xi^{x}=\partial_{\sigma}\xi^{y}=0.
\end{equation}
Hence, the generator of residual symmetries for the line element
\eqref{eq:collgen2} has the general form
\begin{align}
\bm{X}={}&\xi^{\tau}(\tau,\sigma,x,y)\bm{\partial_{\tau}}
       +\xi^{\sigma}(\tau,\sigma,x,y)\bm{\partial_{\sigma}}
 +\xi^{x}(x,y)\bm{\partial_{x}}\nonumber\\
&+\xi^{y}(x,y)\bm{\partial_{y}}+\eta^{I}(x,y,u^{J})\bm{\partial}_{I}.
\end{align}

The Lie-derivative criterion \eqref{eq:crita} gives the following system of
equations for the components of the previous generator
\begin{widetext} {\small
\begin{subequations}
\begin{align}
2\eta^{Q}+\frac{\eta^{a}+\eta^{b}}{a+b}-2\partial_{\tau}\xi^{\tau}
+(b-a)\partial_{\tau}\xi^{\sigma}&=0,\label{f1}\\
(b-a)\eta^{Q}+\frac{b\eta^{a}-a\eta^{b}}{a+b}+\partial_{\sigma}\xi^{\tau}
-ab\partial_{\tau}\xi^{\sigma}-\frac{1}{2}(b-a)(\partial_{\tau}\xi^{\tau}
+\partial_{\sigma}\xi^{\sigma})&=0,\label{f2}\\
-2ab\eta^{Q}+\frac{b^{2}\eta^{a}+a^{2}\eta^{b}}{a+b}
+2ab\partial_{\sigma}\xi^{\sigma}
+(b-a)\partial_{\sigma}\xi^{\tau}&=0,\label{f3}\\
(M-N)(2\eta^{Q}-\partial_{\tau}\xi^{\tau}-\partial_{y}\xi^{y})
+\frac{M-N}{a+b}(\eta^{a}+\eta^{b})-\eta^{M}+\eta^{N}
+(b-a)\partial_{y}\xi^{\sigma}+(Mb+Na)\partial_{\tau}\xi^{\sigma}
-2\partial_{y}\xi^{\tau}&=0,\label{f4}\\
(Mb+Na)(2\eta^{Q}-\partial_{\sigma}\xi^{\sigma}-\partial_{y}\xi^{y})
+\frac{M-N}{a+b}(b\eta^{a}-a\eta^{b})-b\eta^{M}-a\eta^{N}
-2ab\partial_{y}\xi^{\sigma}-(b-a)\partial_{y}\xi^{\tau}
+(M-N)\partial_{\sigma}\xi^{\tau}&=0,\label{f5}\\
\eta^{P}+\eta^{Q}-\partial_{x}\xi^{x}&=0,\label{f6}\\
-2(a+b)\left(e^{-2P}+\frac{MN}{a+b}\right)(\eta^{Q}-\partial_{y}\xi^{y})
-\frac{MN}{a+b}(\eta^{a}+\eta^{b})+N\eta^{M}+M\eta^{N}-2(a+b)e^{-2P}\eta^{P}
-(M-N)\partial_{y}\xi^{\tau}&\nonumber\\
{}+(Mb+Na)\partial_{y}\xi^{\sigma}&=0,\frac{}{}\label{f7}\\
(b-a)\partial_{x}\xi^{\sigma}-(M-N)\partial_{x}\xi^{y}
-2\partial_{x}\xi^{\tau}&=0,\frac{}{}\label{f8}\\
(b-a)\partial_{x}\xi^{\tau}+(Mb+Na)\partial_{x}\xi^{y}
+2ab\partial_{x}\xi^{\sigma}&=0,\frac{}{}\label{f9}\\
-(M-N)\partial_{x}\xi^{\tau}+(Mb+Na)\partial_{x}\xi^{\sigma}
+2(a+b)e^{-P}(\partial_{y}\xi^{x}+\partial_{x}\xi^{y})
+2MN\partial_{x}\xi^{y}&=0.\frac{}{}\label{f10}
\end{align}
\end{subequations}}
\end{widetext}

Since the structural functions of the metric and the spacetime coordinates
are understood as independent variables in jet space, and the
spacetime components of the generator only depend on the later, then
Eqs.~\eqref{f8}-\eqref{f10} are only fulfilled if
\begin{equation}
\partial_{x}\xi^{\tau}=\partial_{x}\xi^{\sigma}=
\partial_{y}\xi^{x}=\partial_{x}\xi^{y}=0.
\end{equation}
Therefore, $\xi^{\tau}$ and $\xi^{\sigma}$ are independent of $x$ while
$\xi^{y}$ and $\xi^{x}$ only are functions of $y$ and $x$, respectively.
Using the same argument together with the complementarity conditions
\eqref{eq:cccolgen}, after taking the derivatives of
Eqs.~\eqref{f1}-\eqref{f7} with respect to $\tau$ and $\sigma$, results in the
additional constraints
\begin{align}
\partial^{2}_{\tau\tau}\xi^{\tau}&=\partial^{2}_{\sigma\sigma}\xi^{\tau}=
\partial^{2}_{\tau\sigma}\xi^{\tau}=\partial^{2}_{\tau{y}}\xi^{\tau}=
\partial^{2}_{\sigma{y}}\xi^{\tau}=0\nonumber\\
\partial^{2}_{\tau\tau}\xi^{\sigma}&=\partial^{2}_{\sigma\sigma}\xi^{\sigma}=
\partial^{2}_{\tau\sigma}\xi^{\sigma}=\partial^{2}_{\tau{y}}\xi^{\sigma}=
\partial^{2}_{\sigma{y}}\xi^{\sigma}=0.\label{f19a}
\end{align}
They imply that the components $\xi^{\tau}$ and $\xi^{\sigma}$ are separable
in sum in all their dependencies and are additionally linear in the Killing
coordinates $\tau$ and $\sigma$
\begin{align}
\xi^{\tau}&=C_{1}\tau+C_{2}\sigma+F_{1}(y),\label{f24}\\
\xi^{\sigma}&=C_{3}\tau+C_{4}\sigma+F_{2}(y).\label{f25}
\end{align}
When we substitute the above solutions in the seventh equations
\eqref{f1}-\eqref{f7} we obtain a linear system of equations for the six
$\eta^{I}$'s plus a constraint. The constraint is just the condition
\begin{equation}
\partial_{x}\xi^{x}=\partial_{y}\xi^{y},
\end{equation}
which is solved as
\begin{align}
\xi^{x}=C_{5}x+C_6,\label{xix}\\
\xi^{y}=C_{5}y+C_7.\label{xiy}
\end{align}
Now Eqs.~\eqref{f1}-\eqref{f6} are a consistent linear system whose solution
is given by
\begin{align}
\eta^{a}&=C_3a^{2}+(C_1-C_4)a-C_2,\label{etaa}\\
\eta^{b}&=-C_3b^{2}+(C_1-C_4)b+C_2,\label{etab}\\
\eta^{P}&=C_5-\frac{1}{2}(C_1+C_4),\label{etap}\\
\eta^{Q}&=\frac{1}{2}(C_1+C_4),\label{etaq}\\
\eta^{M}&=(C_1-C_5)M+C_3aM-\partial_{y}F_2a-\partial_{y}F_1,\label{etam}\\
\eta^{N}&=(C_1-C_5)N-C_3bN-\partial_{y}F_2b+\partial_{y}F_1.\label{etan}
\end{align}
Therefore, the generator of residual symmetries for the noncircular Collinson
\emph{Ansatz} takes the form
\begin{align}
\bm{X}={}&C_{1}
\left(\tau\bm{\partial_{\tau}}+a\bm{\partial_{a}}+b\bm{\partial_{b}}
+M\bm{\partial_{M}}+N\bm{\partial_{N}}\right.\nonumber\\
&-\left.\frac{1}{2}\bm{\partial_{P}}+\frac{1}{2}\bm{\partial_{Q}}\right)
+C_{2}(\sigma\bm{\partial_{\tau}}-\bm{\partial_{a}}+\bm{\partial_{b}})
\nonumber\\
&+C_{3}(\tau\bm{\partial_{\sigma}}+a^{2}\bm{\partial_{a}}
-b^{2}\bm{\partial_{b}}+Ma\bm{\partial_{M}}-Nb\bm{\partial_{N})}\nonumber\\
&+C_{4}\left(\sigma\bm{\partial_{\sigma}}-a\bm{\partial_{a}}-b\bm{\partial_{b}}
-\frac{1}{2}\bm{\partial_{P}}+\frac{1}{2}\bm{\partial_{Q}}\right)\nonumber\\
&+C_{5}(x\bm{\partial_x}+y\bm{\partial_y}-M\bm{\partial_{M}}
-N\bm{\partial_{N}}+\bm{\partial_{P}})\nonumber\\
&+C_6\bm{\partial_x}+C_7\bm{\partial_y}
+F_1\bm{\partial_\tau}-\partial_{y}F_1(\bm{\partial_M}-\bm{\partial_N})
\nonumber\\
&+F_{2}\bm{\partial_{\sigma}}
-\partial_{y}F_{2}(a\bm{\partial_{M}}+b\bm{\partial_{N}}),
\label{eq:gencolgen2}
\end{align}
which is a linear combination of the vector fields reported in the main text
equations \eqref{eq:genscolgen}.


\end{document}